\documentclass[onecolumn,amsmath,amssymb,12pt,superscriptaddress,nofootinbib]{revtex4-1}
\pdfoutput=1

\usepackage[english]{babel}
\usepackage{amsthm}
\usepackage{amssymb}
\usepackage{amsmath}
\usepackage{amsthm}
\usepackage[]{graphicx}
\usepackage{tensor}
\usepackage{color}
\usepackage{framed}
\usepackage{framed}
\usepackage[bookmarks,linktocpage, colorlinks=true, plainpages = false, citecolor = blue,  linkcolor=blue, urlcolor = blue, filecolor = blue]{hyperref} 
\usepackage{natbib}
\usepackage{dcolumn}
\usepackage{bm}
\usepackage{float}
\usepackage{tikz}
\usetikzlibrary{decorations.pathmorphing, patterns,shapes}

\begin{document}

\allowdisplaybreaks
\begin{titlepage}

\title{
Non-Singular Bounces Catalysed by Dark Energy
}
\author{Sebastian F. Bramberger}
\email{sebastian.bramberger@aei.mpg.de}
\affiliation{Max--Planck--Institute for Gravitational Physics (Albert--Einstein--Institute), 14476 Potsdam, Germany}
\author{Jean-Luc Lehners}
\email{jlehners@aei.mpg.de}
\affiliation{Max--Planck--Institute for Gravitational Physics (Albert--Einstein--Institute), 14476 Potsdam, Germany}

\begin{abstract}
\vspace{.5cm}
\noindent 
We investigate classically non-singular bounces caused by dark energy. In the presence of positive spatial curvature, vacuum energy, either in the form of a cosmological constant or a scalar field potential, allows for an open set of initial conditions leading to non-singular bounces, without any violation of the null energy condition. We study anisotropic Bianchi IX cosmologies, and demonstrate that they can even have multiple bounces, accompanied by intricate evolutions of the anisotropies that provide a non-singular analogue of mixmaster crunches. The relation of these solutions to more complete cosmological models, as well as to the recently proposed swampland criteria, are briefly discussed. 
\end{abstract}
\maketitle
\end{titlepage}
\tableofcontents


\section{Introduction}

Within the standard hot big bang model of cosmology, the idea of the big bang is firmly ingrained. This is mostly due to the singularity theorems of Penrose and Hawking \cite{Hawking:1969sw}, which show that under a wide range of conditions, and particularly in the presence of matter that satisfies the null energy condition (as all currently known matter types do), general relativity implies that the current expansion phase must have been preceded by a curvature singularity. The approach to the singularity itself was studied in detail at the classical level by Misner \cite{Misner:1969hg} and also by Belinsky, Khalatnikov and Lifschitz (BKL) \cite{Belinsky:1970ew}. They found the so-called mixmaster behaviour during which the universe shrinks anisotropically - with two spatial directions shrinking while another one expands - in such a way that the total volume becomes smaller and smaller in the approach to the singularity. Which spatial directions shrink and which expand changes ever faster while the universe contracts, leading to chaotic behaviour (see \cite{Damour:2002et} for a review). More realistically however, one would expect quantum gravity, and/or perhaps new types of matter, to become important before the singularity is reached. This new physics may then either explain the creation of spacetime and matter, or their transition from a prior phase of evolution.  

But there remains a rather simple caveat to the singularity theorems: the combination of positive spatial curvature (more specifically, of a compact spatial hypersurface without boundary) and vacuum energy (violating the strong energy condition) allows for classically non-singular bounces to occur, linking a contracting phase of the universe to a subsequent expanding one \cite{Hawking:1967ju}. The simplest and best known example is de Sitter space, in global coordinates. Here we will study similar solutions in more generality, including the effects of anisotropies. As we will discuss, the boundary conditions required for the occurrence of such solutions are rather non-generic, but they do constitute an open set. It currently remains unclear what the eventual importance of these bounces will be for the description of our universe. But importantly, they occur in the presence of known physics, without any exotic matter, without modifications to general relativity and without the need for quantum theory\footnote{See e.g. \cite{Qiu:2011cy,Easson:2011zy,Ijjas:2016tpn,Magueijo:2012ug,Farnsworth:2017wzr} for a small sample of alternative approaches. Skyrmions have also been used to construct non-singular bounce solutions, see e.g. \cite{Ayon-Beato:2015eca,Canfora:2017gno}.}. From this point of view they seem well worth studying. What is more, these non-singular bounces, which also exist in the presence of (sufficiently small) anisotropies, display interesting properties. For instance, they exhibit features reminiscent of the chaotic mixmaster behaviour: multi-bounce solutions exist, with numerous accompanying switches in the expansion rates of different spatial directions. In fact, there does not seem to be any limit to the number of possible bounces, separated by momentary maxima of the size of the universe. We thus find a gradual interpolation between the isotropic de Sitter solution and fully chaotic BKL/mixmaster crunches. Moreover, bounces occur not only in the presence of a cosmological constant, but also for dark energy modelled by a scalar field potential -- in the latter case the potential can be in agreement with the recently proposed swampland criteria, implying that the bounces also present trustworthy solutions from the string theory point of view.

The plan of our paper is as follows: in section \ref{sec:cc} we will first review the Bianchi IX spacetime, which is the most general homogeneous and anisotropic cosmology. This allows us to exhibit non-singular bounces in the presence of a cosmological constant, which constitutes the simplest example of dark energy. In this setting we can also immediately see the link with chaotic mixmaster behaviour, which grows stronger at the edges of the allowed parameter space for bounces. Some features of the bounces can be understood analytically in a restricted spacetime known as axial Bianchi IX \cite{Taub:1950ez,Newman:1963yy}, with the help of an exact solution that was recently discovered, and which we will discuss in section \ref{sec:exact}. Even more generally, this class of bounces persists when including a scalar field with a potential, which will be the topic of section \ref{sec:scalar}. We conclude with an extended discussion in section \ref{sec:discussion} and exhibit another class of analytic, anisotropic bounces in the appendix. We should also comment on the relation of our paper to previous works: various aspects of the bounces of the type considered here, including existence of solutions, ``likelihood'' of bounces, cosmological perturbations across such bounces (especially about the isotropic de Sitter background), were studied over many years -- for a sample of papers see \cite{Barrow:1980en,Hwang:2001zt,Gordon:2002jw,Allen:2004vz,Deruelle:2004fz,Schmidt:1990yg,Barrow:2017yqt}. What is new in our paper is that we provide a systematic overview of explicit numerical solutions, thereby clearly illustrating the link with BKL/mixmaster crunches, and explaining some aspects of these solutions by making use of an analytic solution for axial Bianchi IX, while also discussing the link with the recent swampland conjectures.


\section{Bounces in the presence of a cosmological constant} \label{sec:cc}

\subsection{Bianchi IX}

In order to allow for both positive spatial curvature and anisotropies, we will consider the Bianchi IX metric. One can think of the spatial part of this metric as an evolving three-sphere with two different squashing parameters, so that it represents an anisotropic generalisation of a closed Robertson-Walker spacetime. An alternative point of view is that Bianchi IX represents a fully non-linear completion of a gravitational wave, again in a closed cosmology. The Bianchi IX metric can be written as \cite{Misner:1969hg}
\begin{align}
ds_{IX}^2 = - dt^2 + \sum_m \left( \frac{l_m(t)}{2} \right)^2 \sigma_m^2\,,
\end{align}
where $\sigma_1 = \sin\psi d\theta - \cos \psi \sin \theta d\varphi$, $\sigma_2 = \cos \psi d\theta + \sin \psi \sin \theta d \varphi$, and $\sigma_3 =  d\psi + \cos\theta d\varphi$ are differential forms on the three sphere with coordinate ranges $0 \leq \psi \leq 4 \pi$, $0 \leq \theta \leq \pi$, and $0 \leq \phi \leq 2 \pi.$ We can re-scale 
\begin{align}
l_1(t) &= a(t) \exp \left(\frac{1}{2}\left(\beta_+(t) + \sqrt{3}\beta_-(t)\right)\right) \\
l_2(t) &= a(t) \exp \left(\frac{1}{2}\left(\beta_+(t) - \sqrt{3}\beta_-(t)\right)\right) \\
l_3(t) &= a(t) \exp \left(-\beta_+(t)\right)
\end{align}
such that $a$ represents the average spatial volume while the $\beta$s quantify the deformations of the sphere. When $\beta_- = \beta_+ = 0$ one recovers the isotropic case. 

\begin{figure}[h] 
\begin{center}
\includegraphics[width=0.6\textwidth]{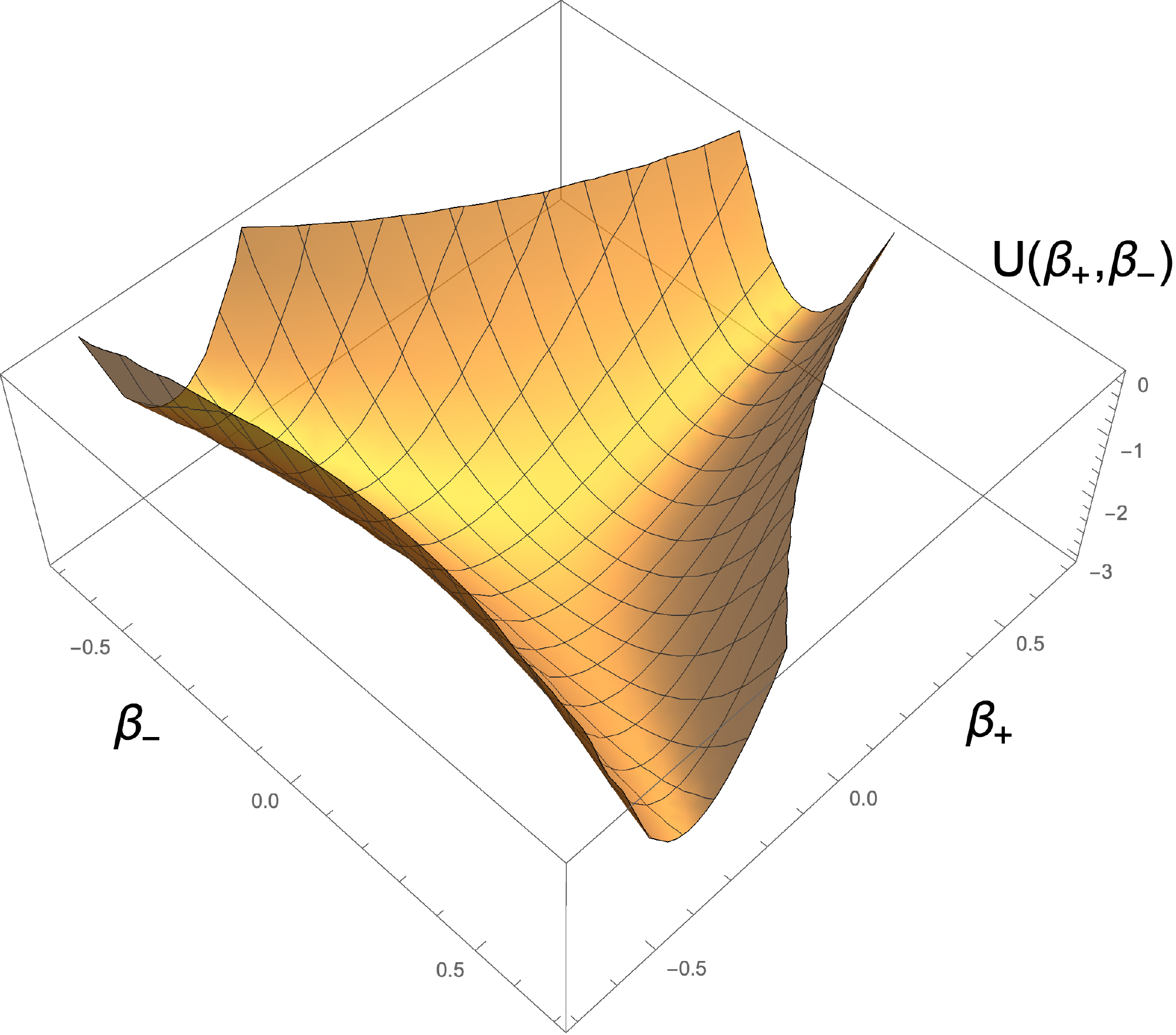}
\caption{The anisotropy potential $U(\beta_+,\beta_-)$. The minimum is at $U(0,0)=-3.$ Around the minimum the potential has an approximate circular symmetry, while at larger values of the anisotropy parameters it has the symmetries of an equilateral triangle. The potential asymptotes to zero from below in the ``corner'' directions. In this plot, only the region where $U<0$ is displayed, for reasons explained in the main text.}
\label{fig:U}
\end{center}
\end{figure}

We will consider general relativity, initially purely in the presence of a constant positive vacuum energy density $\Lambda>0,$ so that in natural units ($8\pi G =1$) the action is
\begin{align}
S = \int d^4x \sqrt{-g} \left( \frac{R}{2} - \Lambda \right)\,.
\end{align}
In the Bianchi IX case, the action reduces to
\begin{align}
S = 2\pi^2 \int dt  a \left[  -3\dot{a}^2  + \frac{3}{4}a^2(\dot{\beta}^2_+ + \dot{\beta}^2_-)   - \left( a^2  \Lambda + U(\beta_+, \beta_-)\right)\right]\,,
\end{align}
where we can see that the anisotropy parameters $\beta_\pm$ have effective kinetic terms analogous to those of scalar fields, while they evolve in the effective potential $U/a^2$ where
\begin{align} \label{anisotropypotential}
U(\beta_+, \beta_-)  & = - 2 \left( e^{ 2 \beta_+ } + e^{-\beta_+ - \sqrt{3}\beta_-} + e^{-\beta_+ + \sqrt{3}\beta_-} \right) + \left( e^{ -4 \beta_+ } + e^{2\beta_+ - 2\sqrt{3}\beta_-} + e^{2\beta_+ + 2\sqrt{3}\beta_-} \right) \\ & = -3 + 6 \left( \dot\beta_+^2 + \dot\beta_-^2 \right) + {\cal{O}}(\beta_\pm^3)\,.
\end{align}
This potential is shown in Fig. \ref{fig:U}. It has a circular symmetry near its minimum at $U(0,0)=-3,$ while at larger values of the anisotropy parameters this symmetry goes over into the symmetry structure of an equilateral triangle. For large anisotropies, the potential ``walls'' rise exponentially fast, and the BKL/mixmaster behaviour alluded to in the introduction effectively corresponds to the anisotropy parameters reflecting off these potential walls, which they do increasingly fast in a contracting universe.

The constraint (Friedman) equation following from the action is given by
\begin{align} \label{Friedman}
3H^2 = \frac{3}{4} \left( \dot{\beta}^2_+ + \dot{\beta}^2_- \right)  + \frac{1}{a^2}U(\beta_+, \beta_-) +   \Lambda \,,
\end{align}
where the expansion/contraction rate is defined as usual as $H\equiv \dot{a}/a,$ while the equations of motion are (using the Friedman equation to simplify the acceleration equation)
\begin{align}
& \frac{\ddot{a}}{a} + \frac{1}{2}\left( \dot{\beta}^2_+ + \dot{\beta}^2_- \right)  -\frac{ \Lambda}{3}  = 0\,, \label{acceleration} \\
& \ddot{\beta}_+ + 3H\dot{\beta}_+  + \frac{2}{3 a^2} U_{,\beta_+} = 0\,, \label{betap} \\
&  \ddot{\beta}_- + 3H\dot{\beta}_-  + \frac{2}{3 a^2} U_{,\beta_-} = 0\,. \label{betam}
\end{align}


\subsection{Time symmetric bounces} \label{sec:sym}

The requirements for a non-singular bounce are straightforward to derive: the equations of motion must allow for the scale factor of the universe to turn around (i.e. they must allow for $\dot{a}=0$) and they must allow for this moment to represent a minimum size, $\ddot{a} > 0.$ At the bounce ($a\equiv a_b, H=0$), the Friedman equation \eqref{Friedman} reads
\begin{align} \label{Friedman2}
-\frac{1}{a_b^2}U(\beta_+, \beta_-) = \frac{3}{4} \left( \dot{\beta}^2_+ + \dot{\beta}^2_- \right)  +  \Lambda \quad \mid_{bounce}\,.
\end{align}
Since the right hand side is positive definite, we see that the anisotropy potential must be negative at the bounce, $U<0,$ which implies that at the bounce, the anisotropy parameters $\beta_\pm$ must reside in the approximately triangular region shown in Fig. \ref{fig:U}.The bounce radius $a_b$ is then given by
\begin{equation}
a_b = \sqrt{\frac{-U}{\Lambda + \frac{3}{4}(\dot\beta_+^2 + \dot\beta_-^2)}}\,.
\end{equation}
Negative $U$ is a necessary condition for a bounce, but it is not sufficient: the acceleration equation \eqref{acceleration} shows that in order to obtain $\ddot{a}>0,$ we must have 
\begin{align}
\frac{3}{2}(\dot\beta_+^2 + \dot\beta_-^2) < \Lambda \quad \mid_{bounce}\,. \label{bounce2}
\end{align} 
Thus we must have suitably small velocities for the anisotropies at the time of the bounce. In other words, for $\dot\beta_\pm  \mid_{bounce}= 0$ we obtain the largest possible set of anisotropy values leading to a bounce. Roughly speaking, the conditions for a successful bounce are that at the bounce the kinetic energy associated with the anisotropies is smaller than the vacuum energy, which in turn must be smaller in magnitude than the (negative) potential energy due to spatial curvature. Note that it is indeed the combination of spatial curvature (leading to $U<0$) and vacuum energy, as exemplified in Eq. \eqref{bounce2}, that allows for non-singular bounces to occur.

\begin{figure}[H] 
\begin{center}
\includegraphics[width=0.45\textwidth]{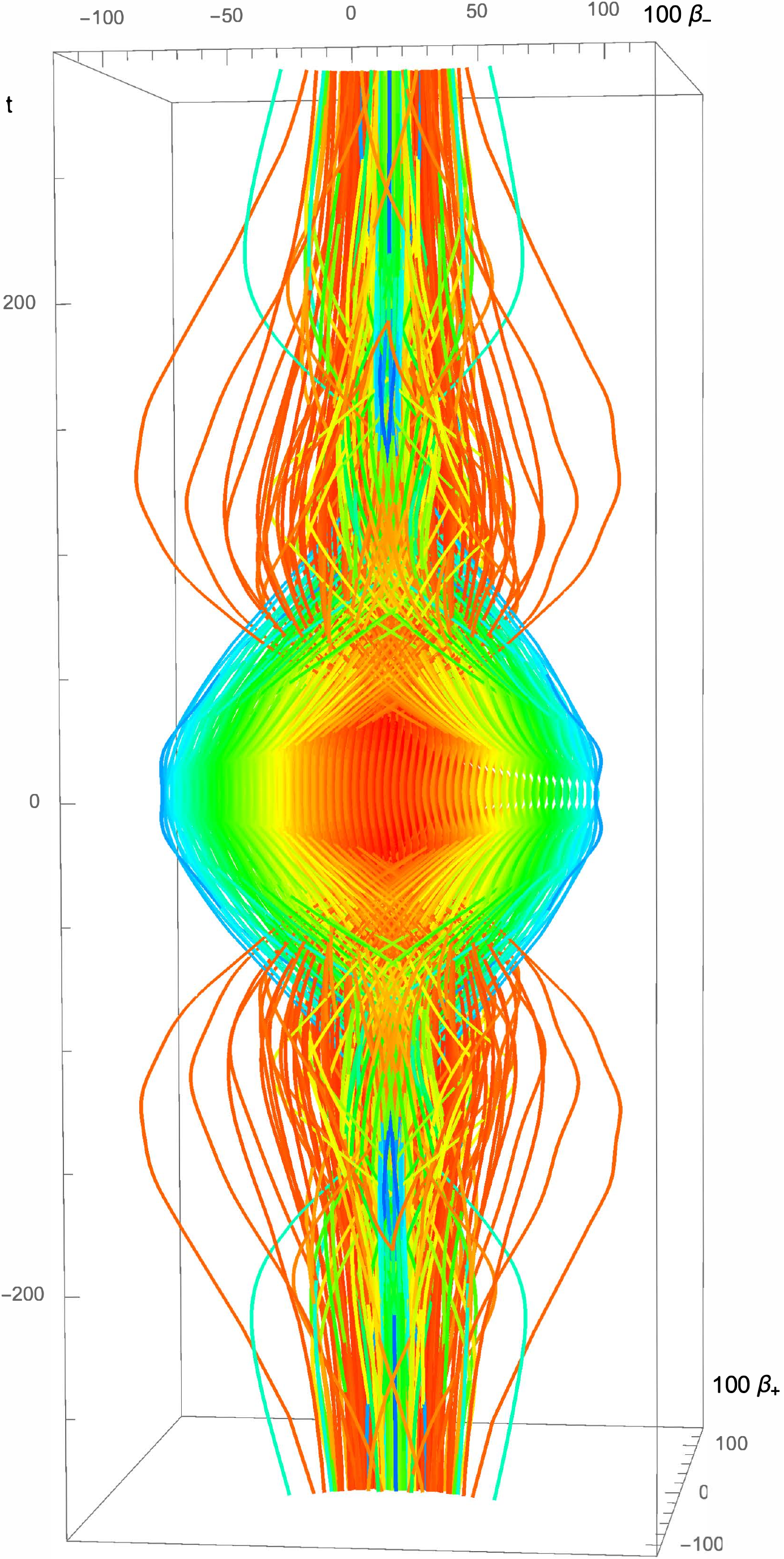}
\caption{This plot shows the evolution of the anisotropies $\beta_\pm$ as a function of time. Time is height in the graph, the plotted ranges are $-7/10< \beta_+ < 11/10, -9/10 < \beta_- < 9/10$ and $-300 < t < 300$, for $\Lambda = 3 \cdot 10^{-4}$ so that the Hubble radius is $1/H=100$ Planck lengths. The bounce occurs in the middle, with zero derivatives $\dot{a}=\dot\beta_+=\dot\beta_-=0$ at $t=0.$ There is a general focussing towards smaller values of the anisotropies away from the bounce. The coloured curves show the evolution of the anisotropies near the bounce for solutions that evolve to a large universe asymptotically. The colour changes as a function of the distance from the isotropic (pure de Sitter) solution located at the centre of the plot, see also the next figures and the text for more details.}
\label{fig:bounce}
\end{center}
\end{figure}

\begin{figure}[H] 
\begin{center}
\includegraphics[width=0.45\textwidth]{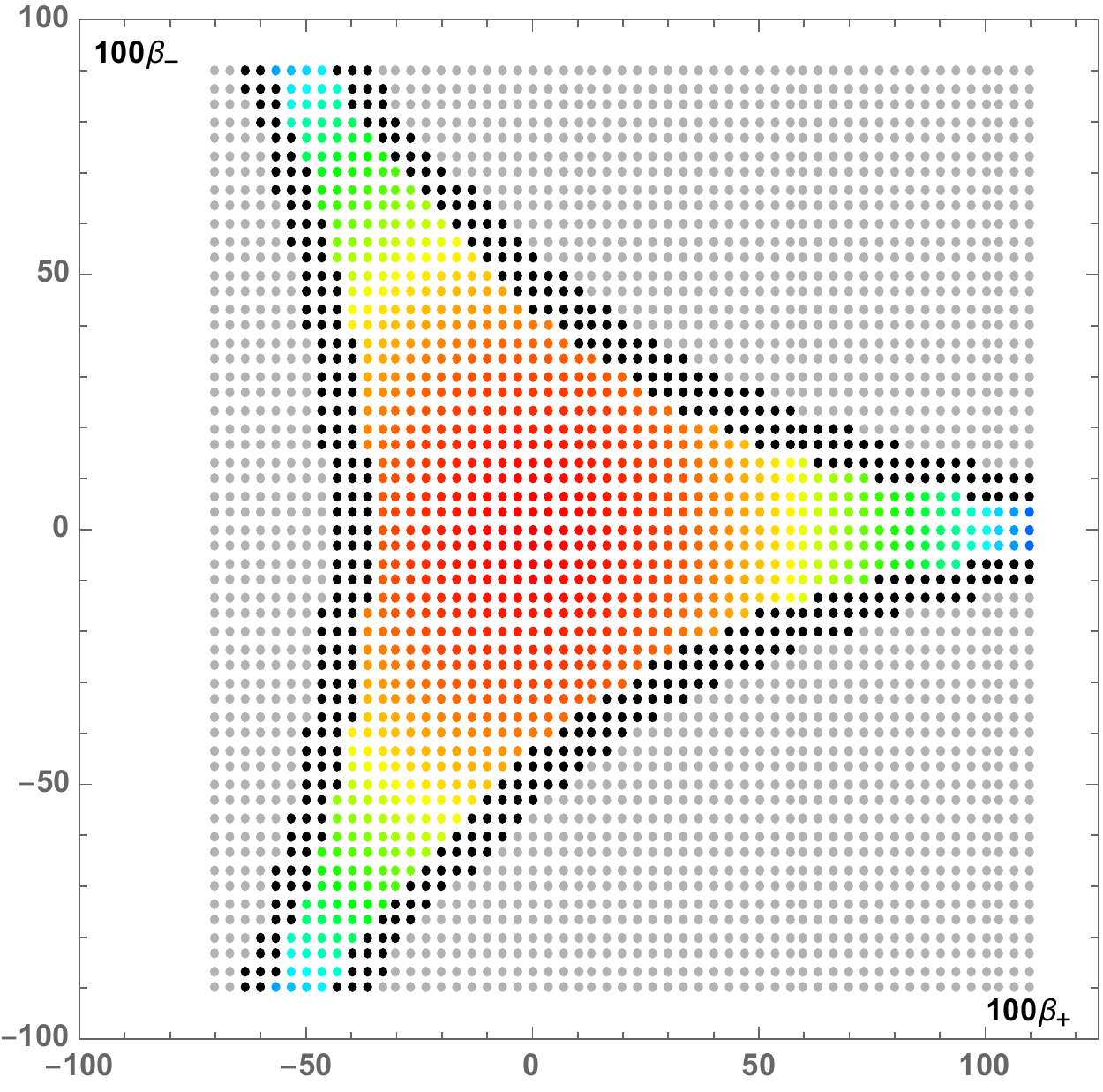}
\includegraphics[width=0.45\textwidth]{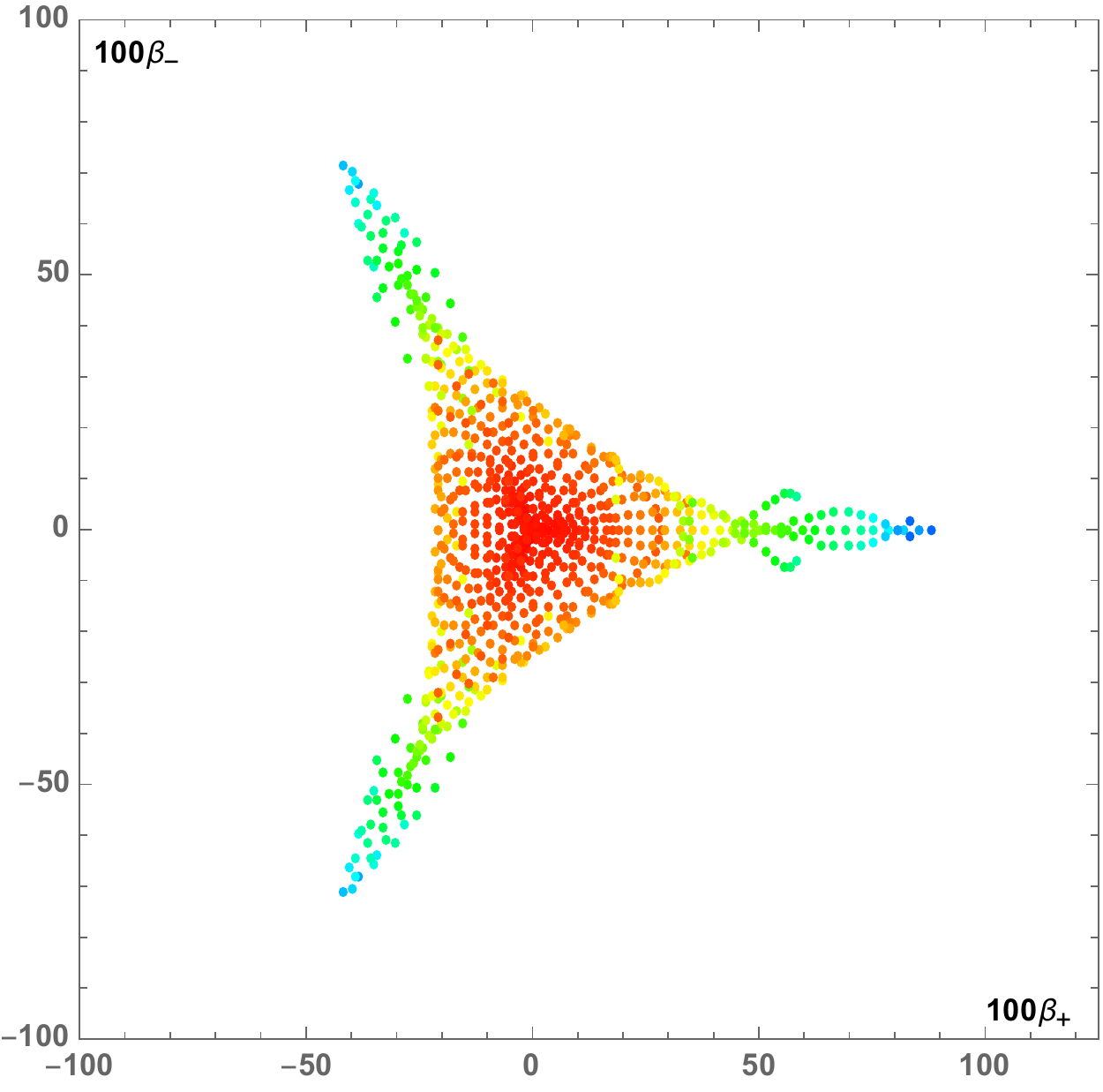} \\
\includegraphics[width=0.45\textwidth]{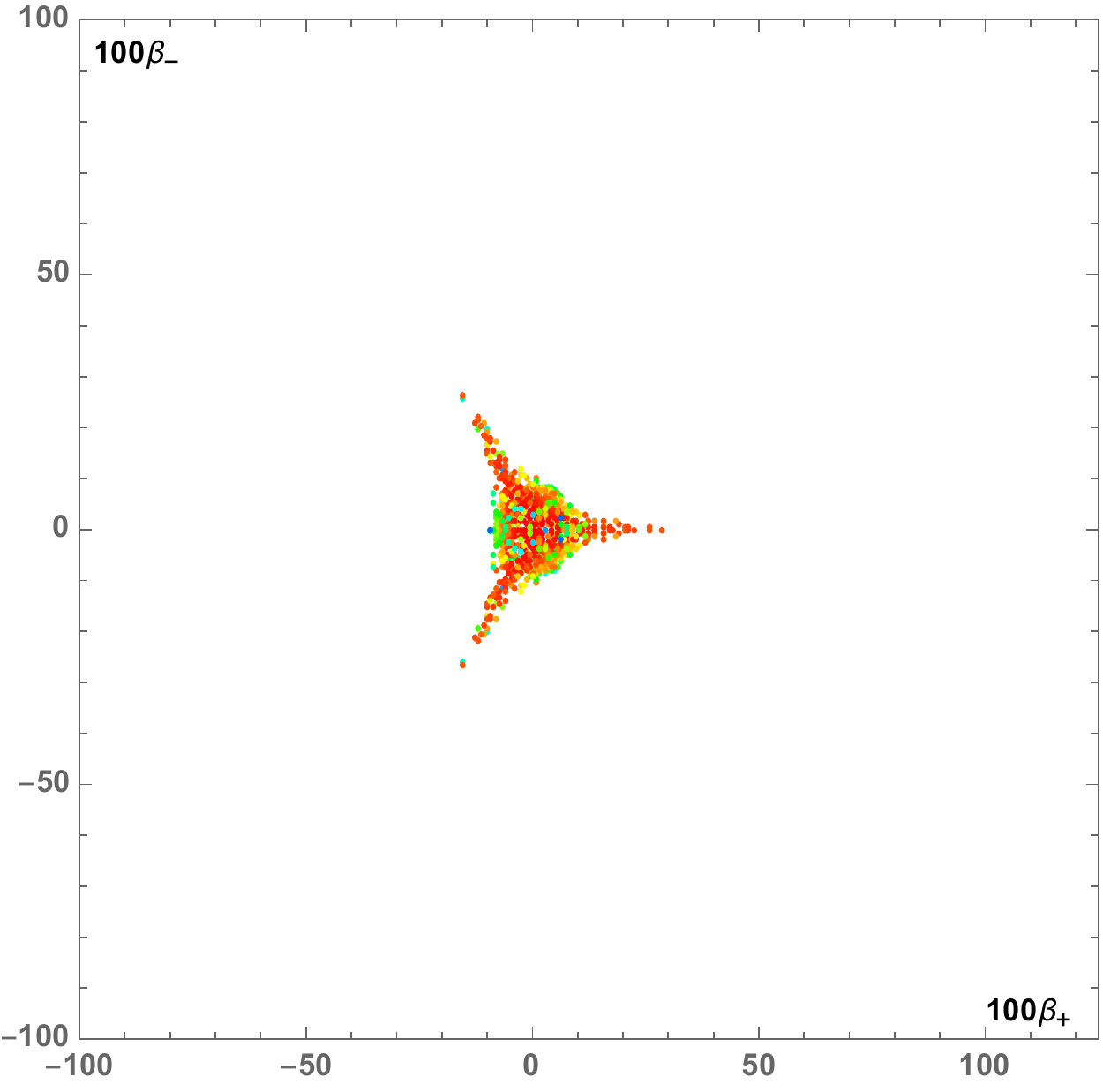}
\includegraphics[width=0.45\textwidth]{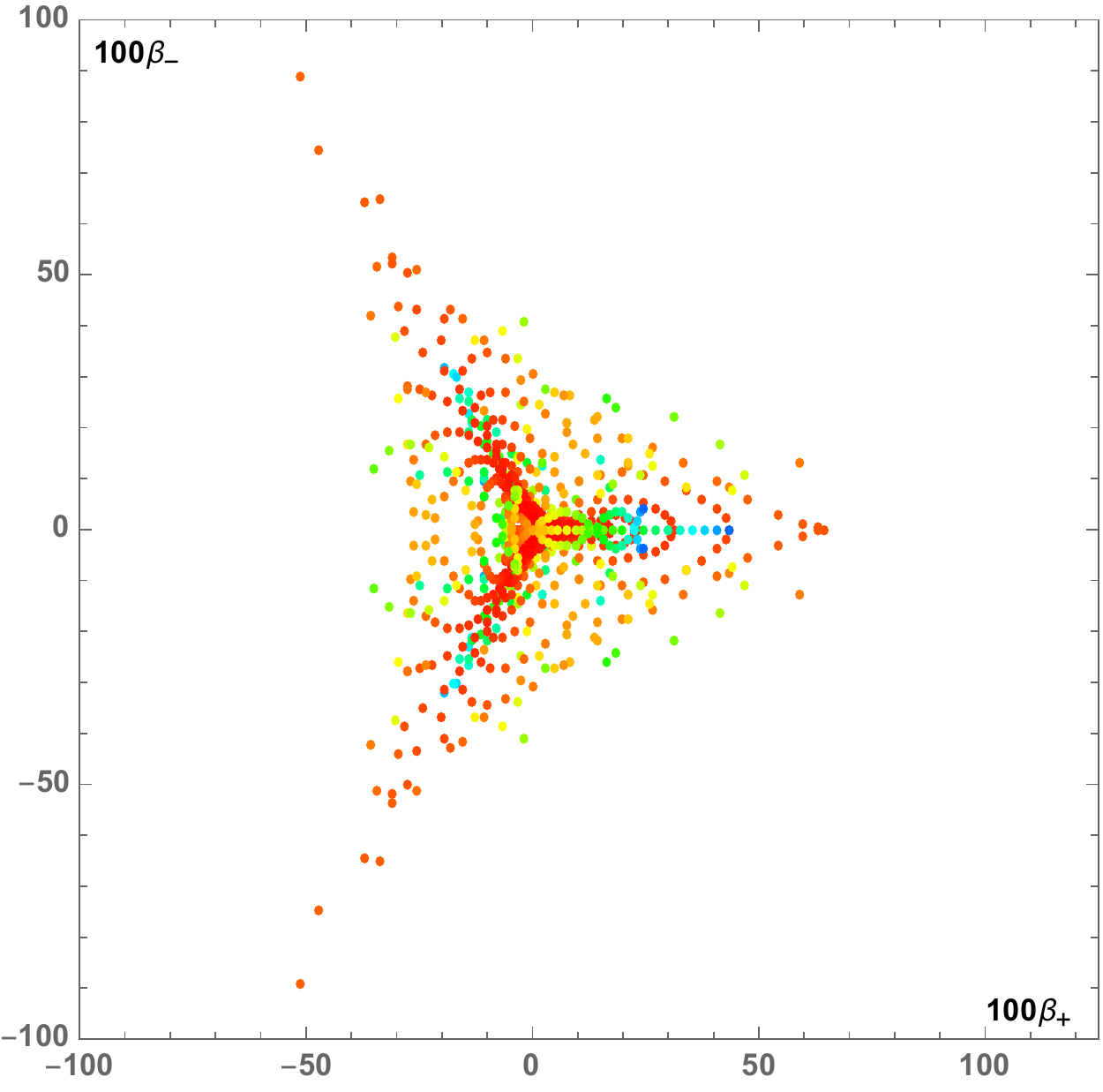}
\caption{Time slices through the previous figure: clockwise from top left at $t=0,50,100,500.$ The anisotropy parameters are re-scaled by a factor of $100.$ Gray dots in the $t=0$ slice indicate values where the potential $U$ is positive (cf. Fig. \ref{fig:U}), and where no bounce can occur. Black dots mark the anisotropy values for which the bounce is followed by a rapid re-collapse. As one can see, the rapid re-collapse region surrounds the conditions for a bounce in all anisotropy directions. Overall, the triangular shape of the anisotropy potential is easily recognisable, and the later time slices show how various solutions reflect off the potential walls, while overall there is a general focussing effect towards smaller anisotropy values away from the bounce.}
\label{fig:bounceslices}
\end{center}
\end{figure}

\begin{figure}[h] 
\begin{center}
\includegraphics[width=0.49\textwidth]{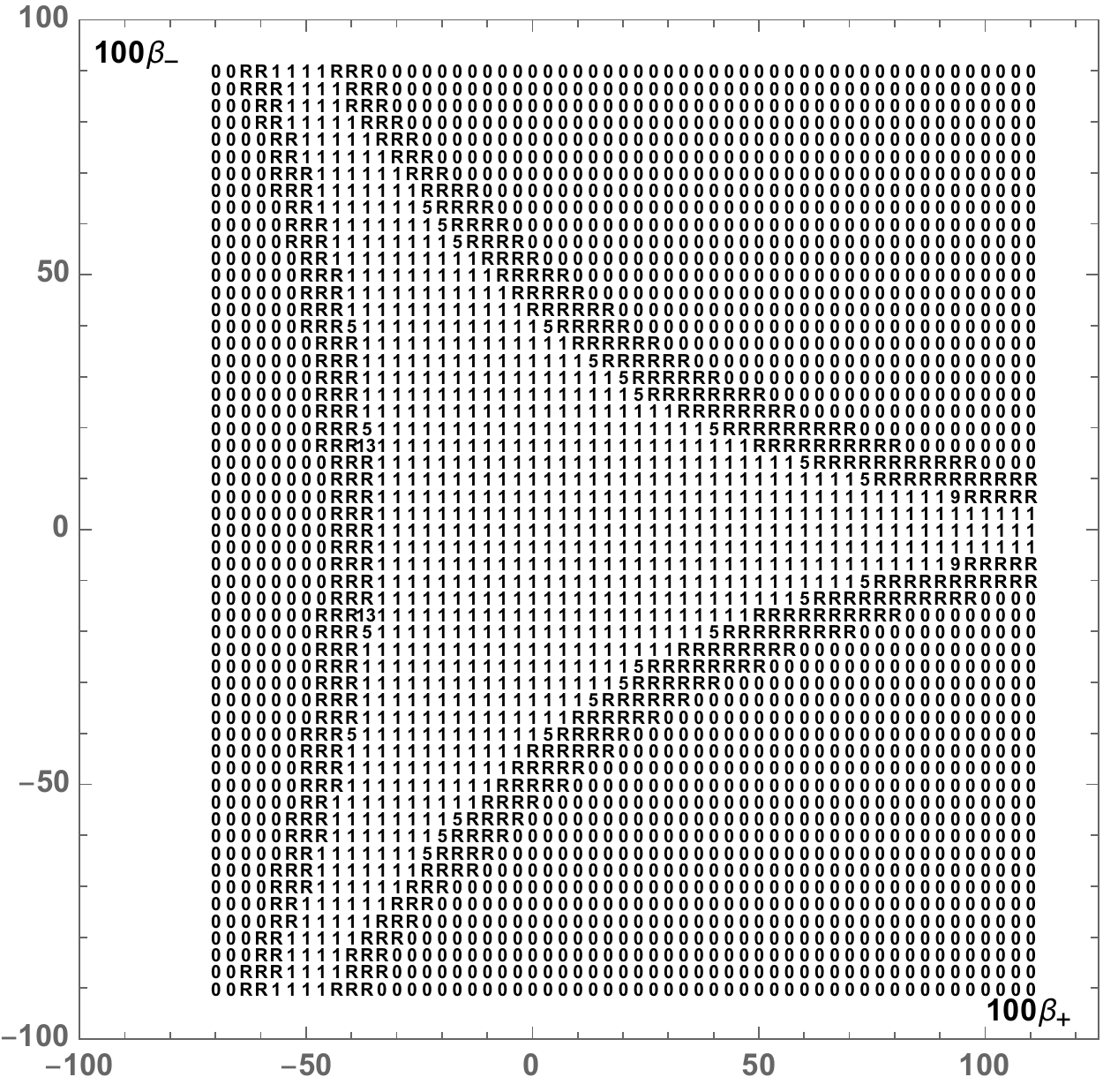}
\includegraphics[width=0.49\textwidth]{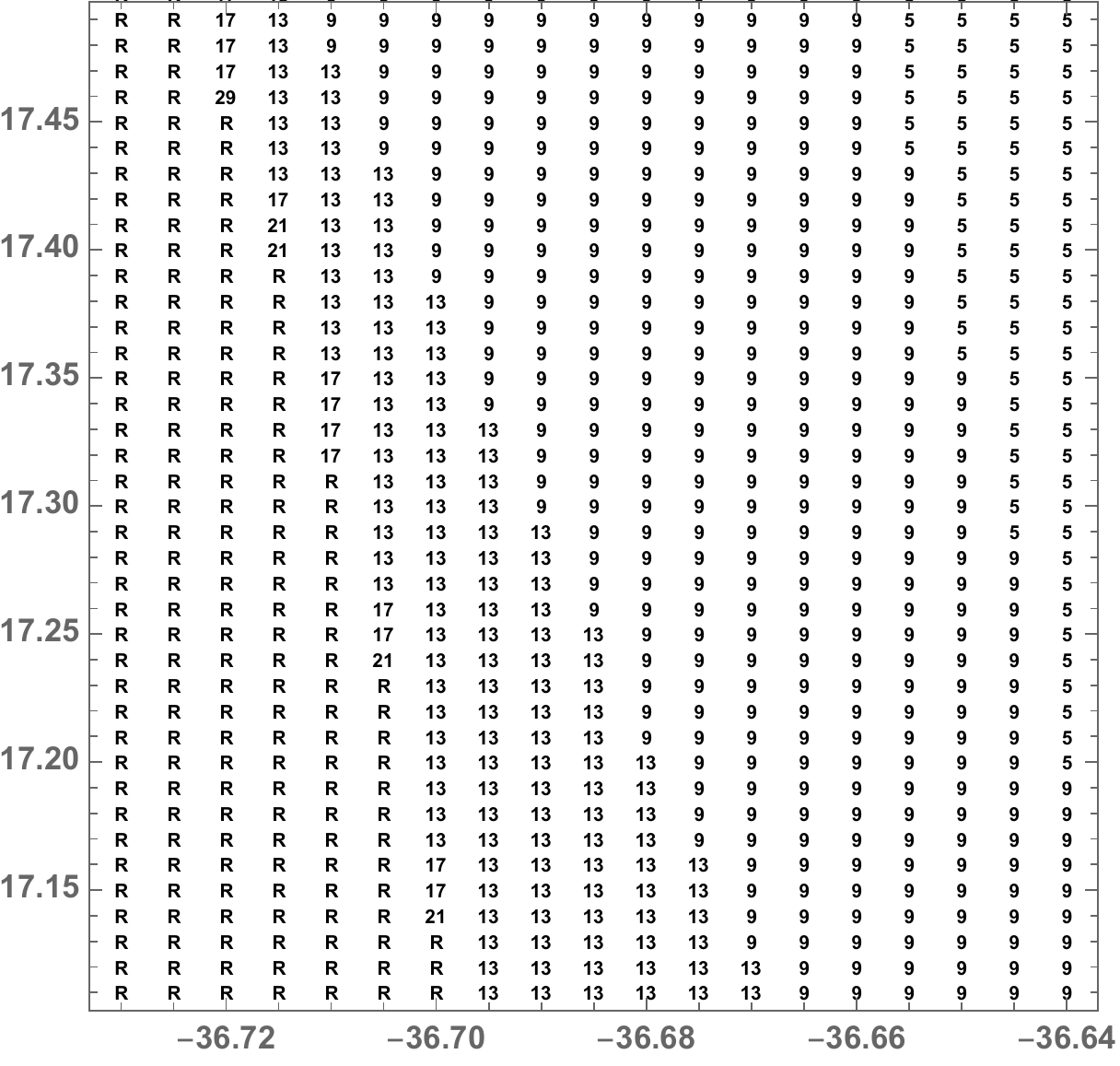}
\caption{These graphs indicate the number of extrema of the scale factor $a(t),$ as a function of the anisotropy values at $t=0$. A value of $1$ corresponds to a single bounce, while $5$ for instance implies three bounces separated by two local maxima of $a$. An $R$ marks a bouncing solution that rapidly re-collapses to a singularity, while $0$ means that no bounce is possible at all. The plot on the right is a zoom-in near the edge of the region of re-collapse.}
\label{fig:bouncenumber}
\end{center}
\end{figure}

\begin{figure}[h] 
\begin{center}
\includegraphics[width=0.45\textwidth]{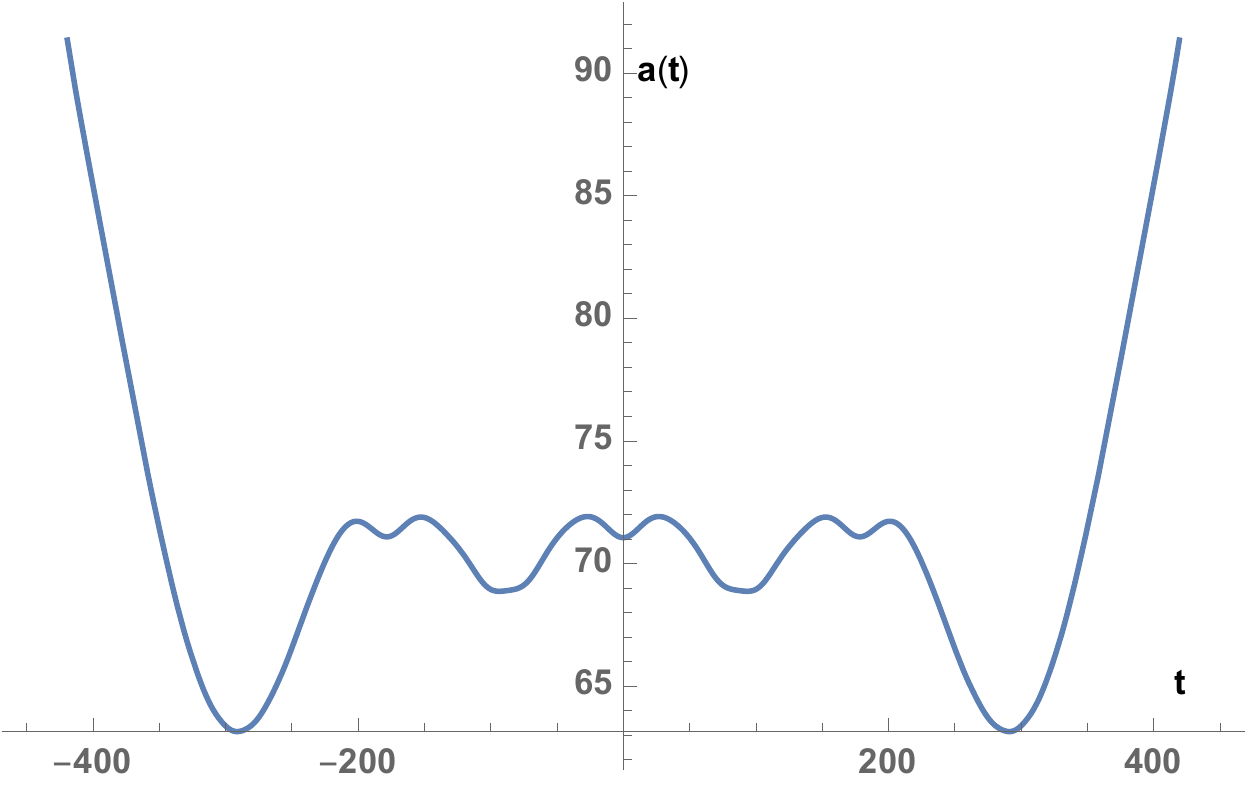}
\includegraphics[width=0.45\textwidth]{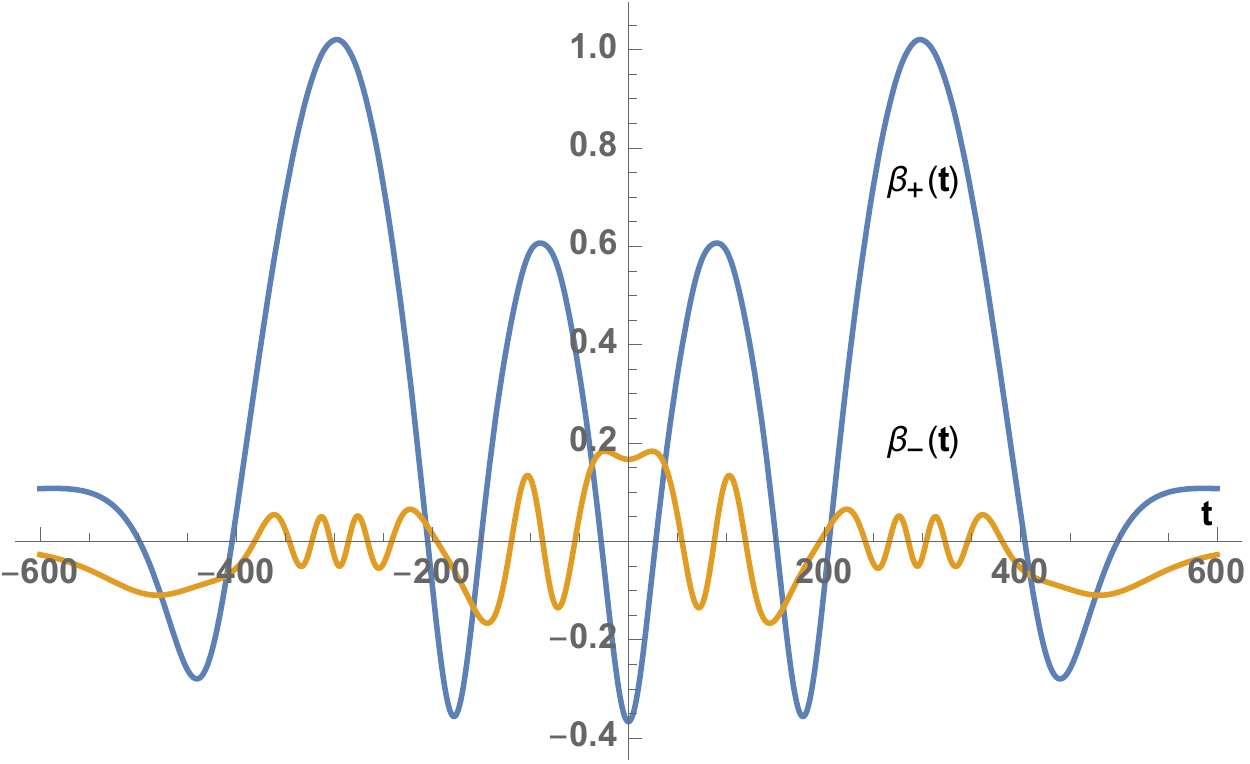}
\caption{These two graphs show the evolution of the scale factor and the anisotropy parameters for the time symmetric solution with $\beta_+=-11/30, \beta_-=1/6$ at the bounce. For this solution there are a total of $7$ bounces occurring in succession, while the anisotropy parameters undergo elaborate reflections off the walls of the anisotropy potential. Multi-bounce solutions such as this one occur near the edges of the allowed parameter space, as evidenced in Fig. \ref{fig:bouncenumber}.}
\label{fig:bounce13}
\end{center}
\end{figure}

With the exception of a special sub-class of solutions presented in subsection \ref{sec:exact} and for which an analytic expression exists, we must find the bouncing solutions numerically. We will start with the best possible case, where we demand that the time derivatives of the anisotropy parameters are set to zero at the moment of the bounce,  $\dot\beta_+(t_b)=\dot\beta_-(t_b)=0$ at $\dot{a}(t_{b})=0.$ Without loss of generality we will choose the origin of the time coordinate to be at the bounce, $t_b=0.$ Since the derivatives are all zero at the bounce, these solutions will be symmetric in time, i.e. the contraction phase leading up to the bounce will be the time reverse of the ensuing expanding phase. Our numerical results for this case are presented in Figs. \ref{fig:bounce} -- \ref{fig:bounce13}. In all these plots we have chosen $\Lambda = 3 \times 10^{-4},$ so that the Hubble radius is given by $1/H=\sqrt{3/\Lambda}=100$ in Planck units, i.e. we made the assumption that the vacuum energy was large in the early universe. The solutions presented here however exist for any chosen value of $\Lambda$ and can be obtained using suitable re-scalings of the coordinates.

Fig. \ref{fig:bounce} shows the evolution of the anisotropy parameters as a function of time. Each trajectory represents a bouncing solution, with the colour determined by the distance $\beta_+^2+\beta_-^2$ in anisotropy space (at $t=0$) from the isotropic de Sitter solution for which $\beta_\pm=0.$ Time slices through these solutions are presented in Fig. \ref{fig:bounceslices} at times $t=0,50,100,500,$ where we should keep in mind that the characteristic time scale implied by the vacuum energy is $100$ Planck times for our choice of $\Lambda.$ Each coloured trajectory describes a successful bouncing solution, in the sense that at large early/late times these solutions contract/expand exponentially.

They may, however, contain short time intervals of re-collapse, followed by another bounce. This is illustrated in Fig. \ref{fig:bouncenumber} where the number of extrema of the scale factor is shown. A value of $1$ implies a standard non-singular bounce solution for which the scale factor has a typical ``U'' shape as a function of time. By contrast, a value of $3$, for instance, implies that there are two bounces separated by a local maximum of the scale factor, i.e. the scale factor has a profile that resembles the letter ``W''. We deem a solution to be an unsuccessful bounce if shortly before/after the bounce the scale factor re-collapses to zero size, leading to a curvature singularity. Such re-collapsing solutions are marked with the letter $R$ in Fig. \ref{fig:bouncenumber}. As we derived above, the anisotropy potential must be negative in order for a bounce to occur. When this is not the case, i.e. when no bounce can occur at all, not even a temporary one followed by re-collapse, we assigned the entry $0$ in Fig. \ref{fig:bouncenumber}. From this graph we can see that the region where bounces occur is separated from the region where they cannot occur by re-collapsing solutions that simply shift the singularity in time, without eliminating it. The edge of the re-collapsing region is formed by what might very well be the most interesting bouncing solutions from a mathematical viewpoint: here there exist solutions with increasing numbers of intermediate bounces, and intricate evolutions of the anisotropy parameters. An example with $13$ extrema of the scale factor, i.e. $7$ bounces and $6$ local maxima of $a,$ is shown in Fig. \ref{fig:bounce13}. The plot of the evolution of the anisotropies shows that this solution repeatedly reflects off the walls of the anisotropy potential $U(\beta_+,\beta_-),$ reminiscent of the BKL/mixmaster behaviour of singular crunches. The evolution here reveals a substantial sensitivity to initial conditions, although it is not chaotic in the BKL sense, in that there are only a finite number of such reflections before a non-singular bounce occurs. Nevertheless, as one approaches the edge of the re-collapse region in ever smaller intervals, there seems to be no limit to the number of bounces, as illustrated by the right panel in Fig. \ref{fig:bouncenumber}. The latter graph for instance includes a solution with $15$ bounces separated by $14$ local maxima of the scale factor. It would be interesting, though computationally intense, to find the shape of the curves delineating the borders between solution regions with different numbers of bounces. This question must, however, be left for future work \footnote{Analyses of the chaotic nature of isotropic solutions (in the presence of a massive scalar field) have already been performed in \cite{Kamenshchik:1998ue,Kamenshchik:1998xx,Kamenshchik:1998ix}.}.

Overall, there is a significant focussing of the anisotropies towards smaller values as one goes away from the bounce. Also, in all successful bounce solutions, the anisotropies rapidly reach approximately constant values at early and late times, with all of the interesting evolution confined to the time period of the bounce.  For the case of zero (or very small) velocities at the bounce, we can also understand the focussing effect analytically. This is because the equations of motion \eqref{betap}, \eqref{betam} for the anisotropies simplify near the bounce to give
\begin{equation}
\ddot\beta_\pm \approx -\frac{2}{3a_b^2}U_{,\beta_\pm}\,.
\end{equation} 
Since the effective potential $U$ rises from the origin in all directions of increasing anisotropy, the above equation implies that the anisotropy will be reduced as we go away from the bounce.

\subsection{Time asymmetric bounces}

\begin{figure}[h] 
\begin{center}
\includegraphics[width=0.45\textwidth]{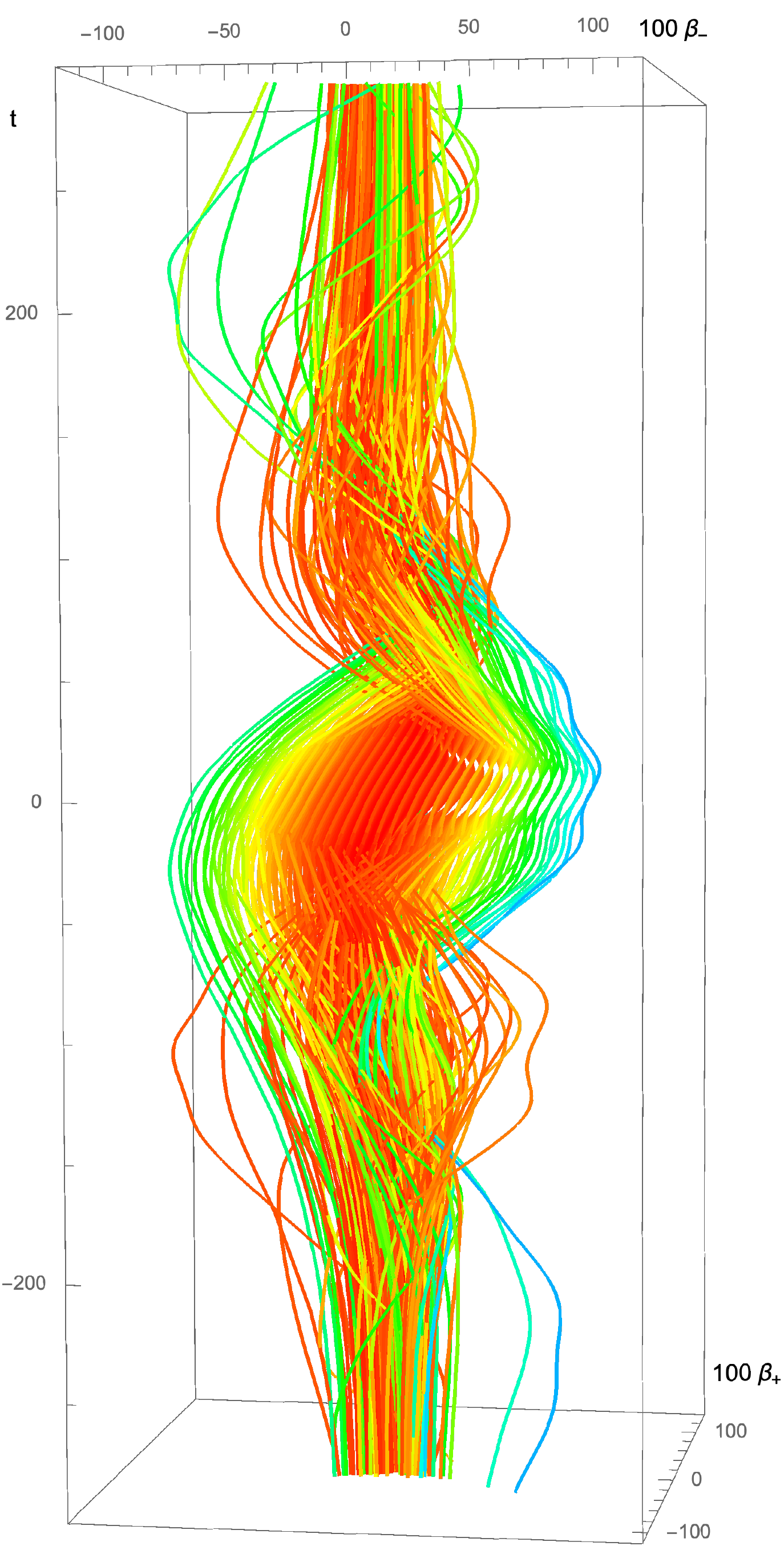}
\includegraphics[width=0.45\textwidth]{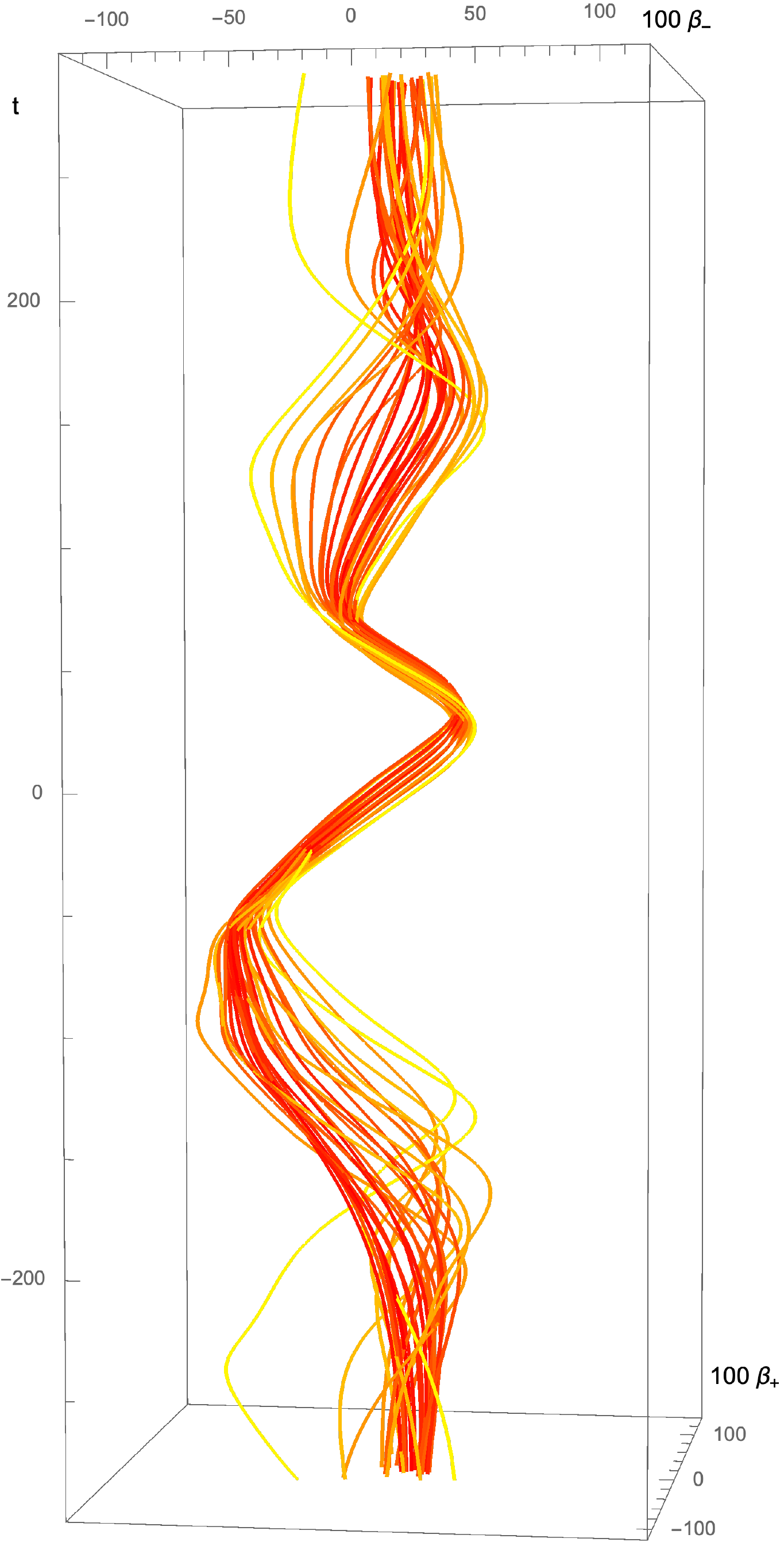}
\caption{This plot shows the evolution of the anisotropies $\beta_\pm$ as a function of time. Time is height in the graph, the plotted ranges are $-7/10< \beta_+ < 11/10, -9/10 < \beta_- < 9/10$ at the bounce, and $-300 < t < 300$, for $\Lambda = 3 \cdot 10^{-4}$ so that the Hubble radius is $1/H=100$ Planck lengths. Initial conditions are imposed at $t=0$, where $\dot{a}=0$ and $\dot\beta_+=\dot\beta_-=1/200$ (left graph) and $\dot\beta_+=\dot\beta_-=1/80$ (right graph).}
\label{fig:bounceasy}
\end{center}
\end{figure}

\begin{figure}[h] 
\begin{center}
\includegraphics[width=0.45\textwidth]{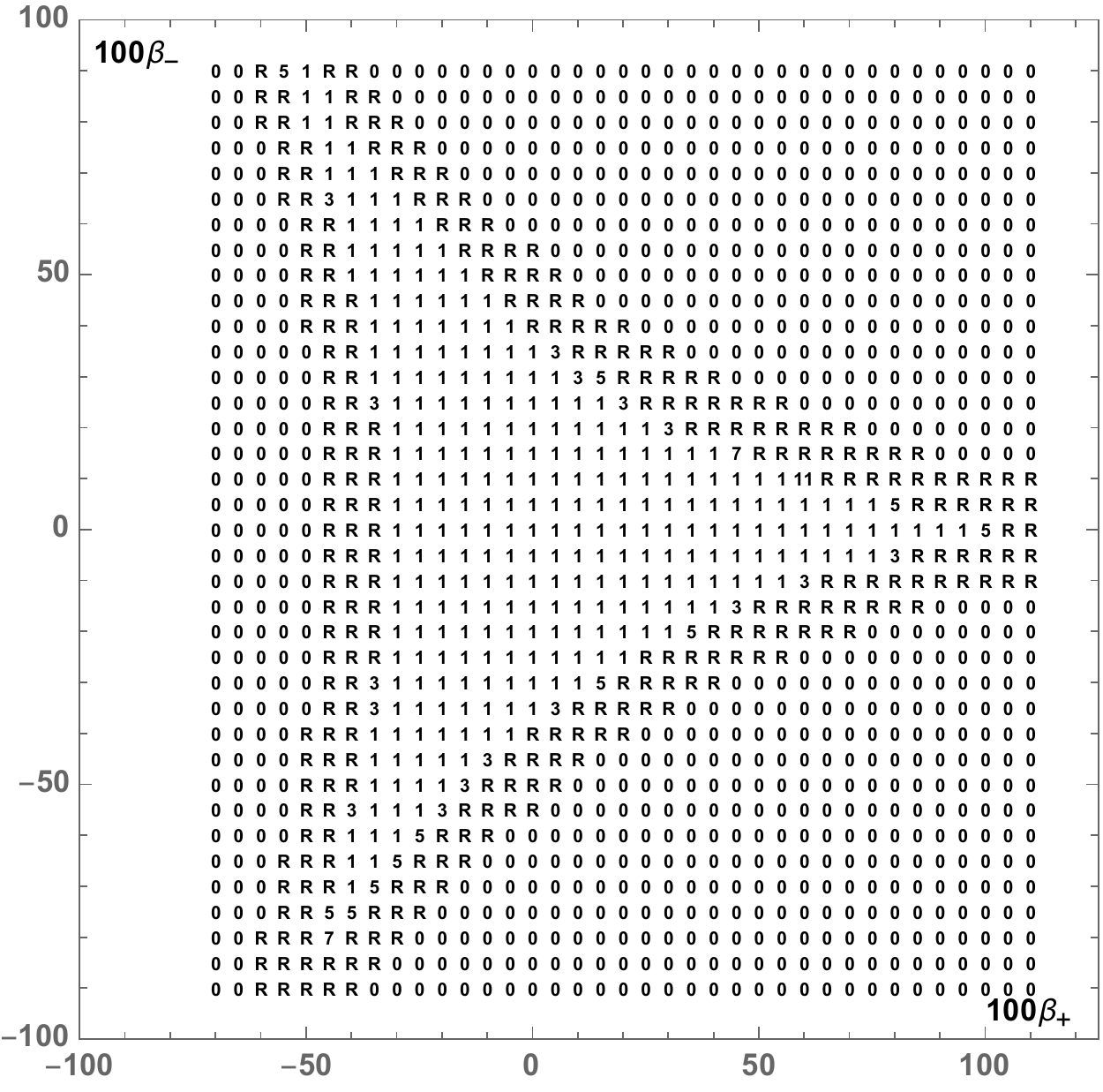}
\includegraphics[width=0.45\textwidth]{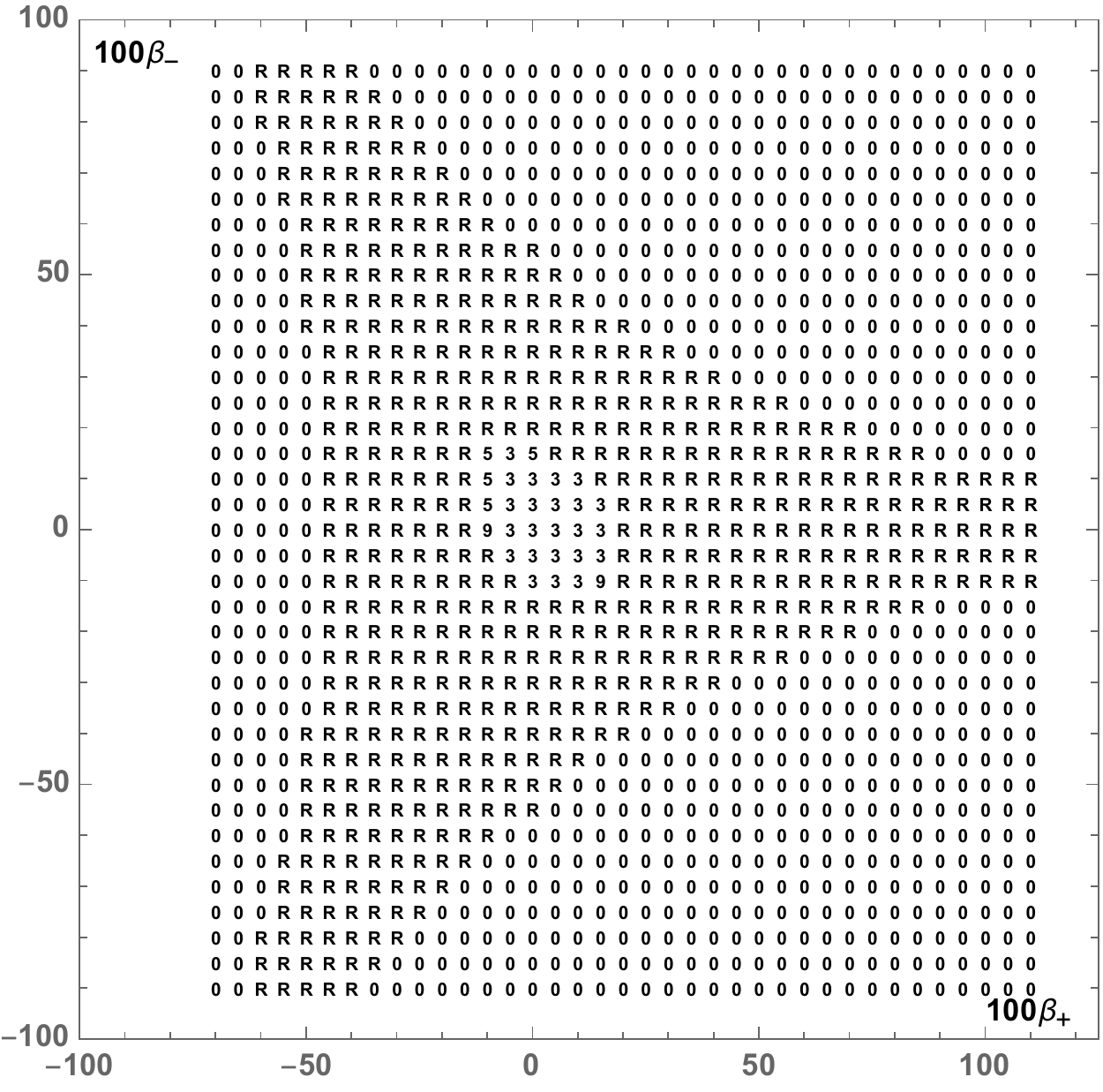}
\caption{The number of extrema of the scale factor $a(t)$ for the solutions plotted in Fig. \ref{fig:bounceasy}.}
\label{fig:bounceasynumber}
\end{center}
\end{figure}

We can now extend these results by allowing for non-zero time derivatives of the anisotropy parameters at the bounce. The allowed range is indicated by the bound in Eq. \eqref{bounce2}, which can be read to say that the ``kinetic energy'' in the anisotropy must be smaller than half of the energy density of vacuum energy. Numerically, we find that, increasing this kinetic energy, the results of the previous section are modified very little until one gets close to the upper bound. The left panel in Fig. \ref{fig:bounceasy} for instance shows the results for the case where we take $\dot\beta_+(0)=\dot\beta_-(0)=1/200,$ implying that $\frac{3}{2}(\dot\beta_+^2+\dot\beta_-^2)=\frac{1}{4}\Lambda$ at $t=0.$ Even for these values which are just a factor of $1/4$ away from the upper bound, the main effects are a slight time asymmetry in the solutions and a modest reduction of the available anisotropy space leading to bounces. The left panel in Fig. \ref{fig:bounceasynumber} illustrates this.

\begin{figure}[h] 
\begin{center}
\includegraphics[width=0.45\textwidth]{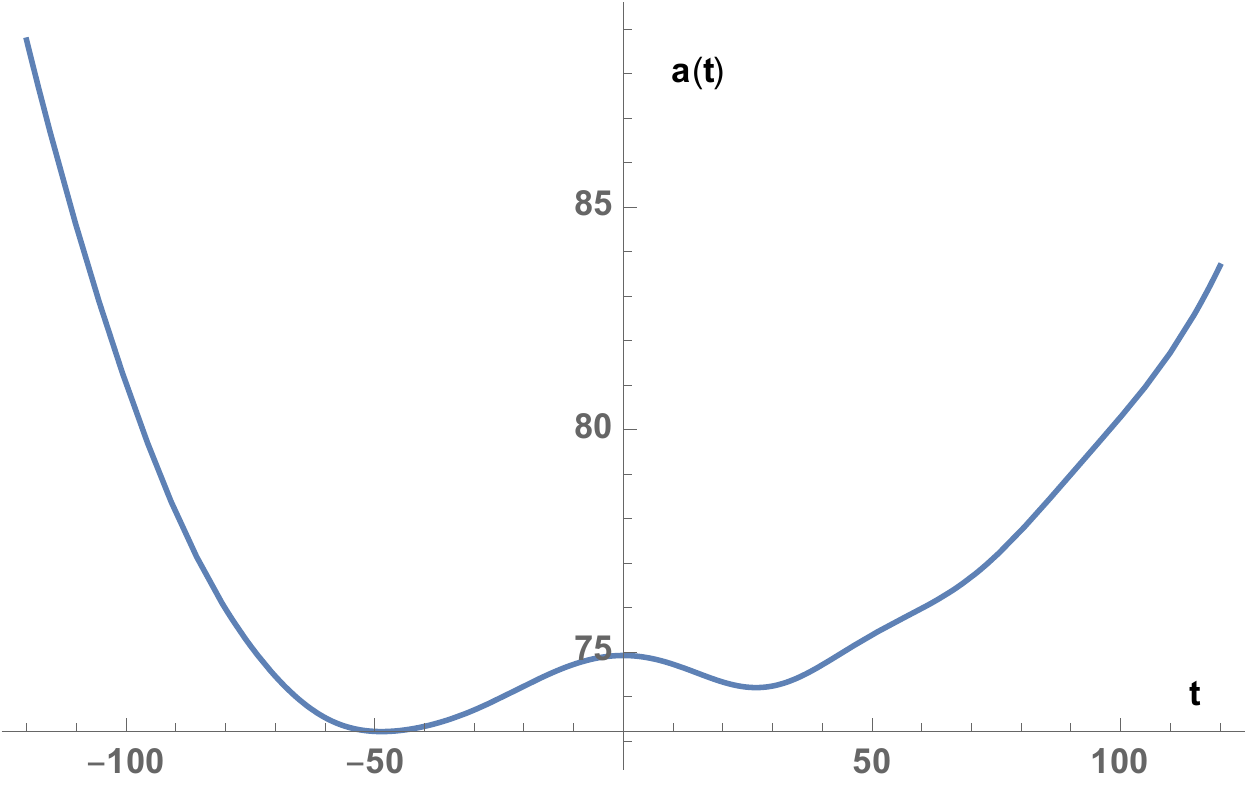}
\includegraphics[width=0.45\textwidth]{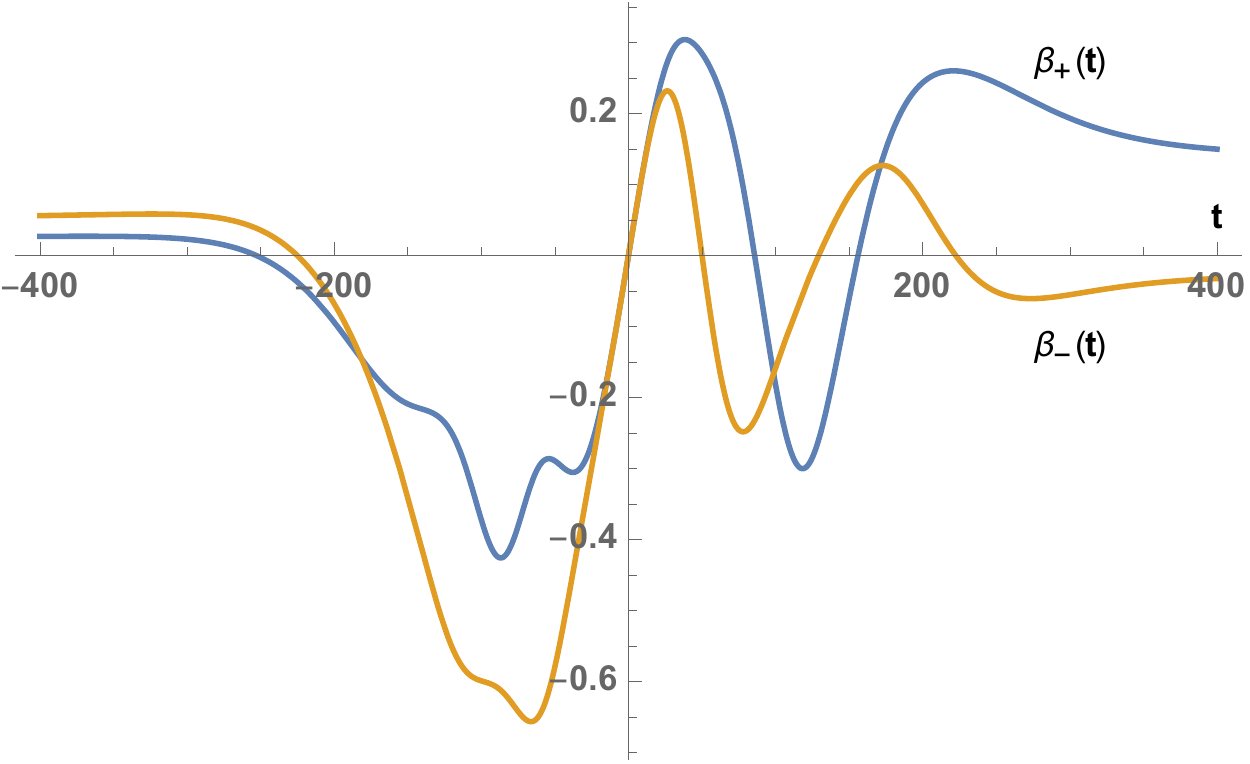}
\caption{These graphs show the evolutions of the scale factor and of the anisotropy parameters as a function of time, for the solution with $\beta_+(0)=\beta_-(0)=0,$ but with large velocities $\dot\beta_+(0)=\dot\beta_-(0)=1/80$. Although the kinetic energy at $t=0$ is larger than the value that could lead to a bounce, away from $t=0$ two bounces nevertheless occur since the energy in the anisotropies is slightly reduced there.}
\label{fig:bounce3}
\end{center}
\end{figure}

Interestingly, one may increase the velocities of the $\beta$ parameters at $t=0$ even slightly beyond the bound of Eq. \eqref{bounce2}, and still obtain non-trivial results -- see the right panels in Figs. \ref{fig:bounceasy} and \ref{fig:bounceasynumber}, where we took $\dot\beta_+(0)=\dot\beta_-(0)=1/80,$ implying that $\frac{3}{2}(\dot\beta_+^2+\dot\beta_-^2)=\frac{25}{16}\Lambda$ at $t=0.$ Simple bounces have now disappeared (in agreement with the derived bound), but multi-bounce solutions may still exist, since the anisotropy parameters may evolve to smaller velocities away from $t=0$ and lead to bounces there. An example of such a solution with $3$ extrema of the scale factor, translating into two bounces separated by one local maximum of $a(t),$ is plotted in Fig. \ref{fig:bounce3}. Overall, the parameter space leading to bounce solutions is drastically reduced when the kinetic energy in the anisotropy is this large. From these considerations it seems clear that non-singular bounces of the type discussed here can only have played a role in the early universe if the vacuum energy was very large., and if the growth of anisotropies during a prior contracting phase was suitably mitigated. We will discuss this aspect in more detail in section \ref{sec:discussion}.

\subsection{Axial Bianchi IX: an exact solution} \label{sec:exact}

Our discussions so far were based on numerical solutions to the equations of motion, for various boundary conditions. In fact it seems difficult to imagine that a general analytic solution can be found for full Bianchi IX non-singular bounces. However, there exists a special subset of solutions for which an exact result may be found. We will present it here, as it confirms our numerical results in the relevant parameter region, and provides useful insights into the general structure of this subset of solutions.

\begin{figure}[h] 
\begin{center}
\includegraphics[width=0.45\textwidth]{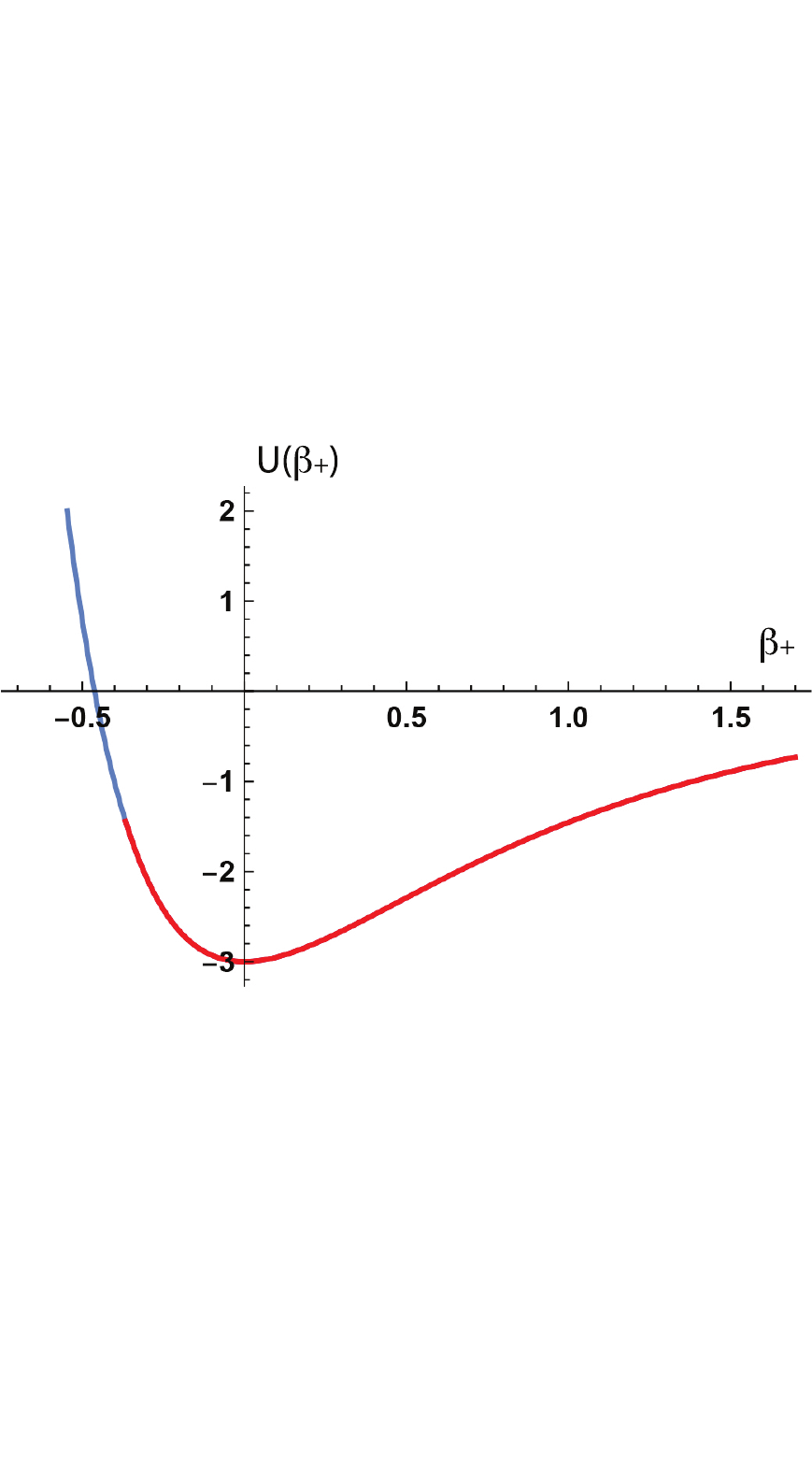}
\includegraphics[width=0.45\textwidth]{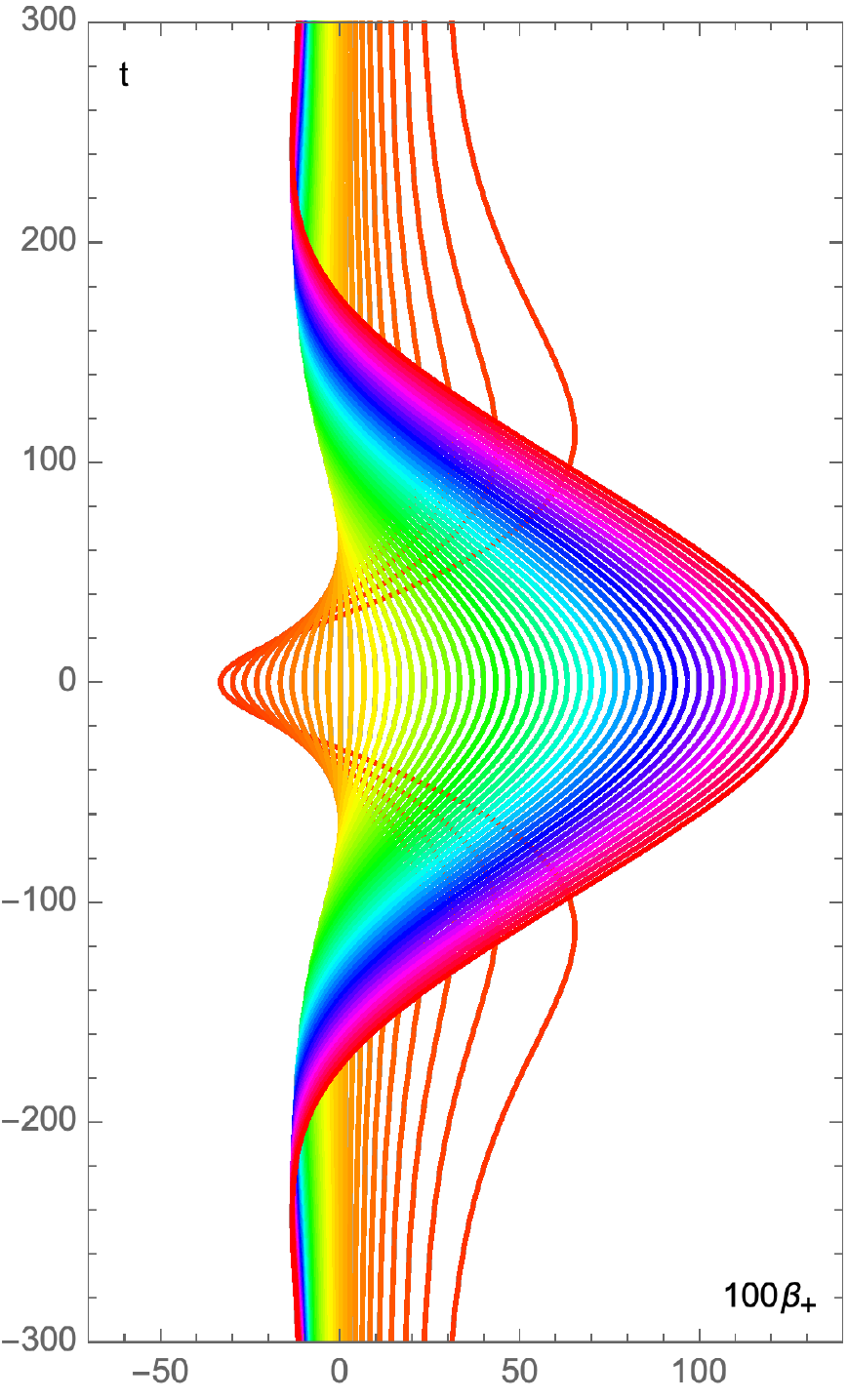}
\caption{The left plot depicts the anisotropy potential $U(\beta_+,\beta_-=0)$ along one of the axes of symmetry, here chosen to be the axis $\beta_-=0$. The range over which non-singular bounces can occur is marked in red. Bounces occur for arbitrarily large positive anisotropies in this direction, and the solutions are strongly focussed towards zero away from the bounce. There is a minimum value $\beta_+ =\frac{1}{3}\ln(\frac{1}{3})\approx -0.366$ below which the solutions rapidly re-collapse. Below $\beta_+=\frac{1}{3}\ln(\frac{1}{4})\approx -0.462$ the anisotropy potential is positive an no bounce can occur at all.}
\label{fig:bounceaxis}
\end{center}
\end{figure}

One may consistently truncate the equations of motion \eqref{Friedman} - \eqref{betam} to a simpler system with just one deformation parameter, along any of the three axes of symmetry of the full Bianchi IX metric. The simplest choice is the axis defined by $\beta_-=0,$ and along this axis $\beta_-$ will then not be sourced by non-zero $\beta_+$. Thus, the anisotropy space is reduced from $2$ to just $1$ dimension, and sometimes this is called the axial Bianchi IX case. With the choice $\beta_-=0$, the effective anisotropy potential simplifies to the form
\begin{align} \label{anisotropycornerpotential}
U(\beta_+, \beta_-=0)  = - 4 e^{-\beta_+} + e^{ -4 \beta_+ } \,.
\end{align}
This potential is shown in the left panel of Fig. \ref{fig:bounceaxis}. It contains a local minimum at negative values of the potential, and asymptotes zero from below as $\beta_+\to \infty.$ The bounce criterium that $U$ must be negative to allow for a non-singular bounce thus suggests that it might be possible to find bounce solutions for arbitrarily large values of $\beta_+$ as long as $\beta_-=0,$ and we will see that this expectation is borne out.

Anabal\'{o}n and Oliva (AO) recently found an exact bouncing cosmology for the axial Bianchi IX system \cite{Anabalon:2018rzq}. Their solution is most easily expressed in a coordinate system different to the one we have been using so far -- more specifically we will take the line element to be
\begin{align}
ds^2 = - \frac{4l^4}{\sigma^2 f(\tau)}d\tau^2 + g(\tau)(\sigma_1^2+\sigma_2^2) + f(\tau) \sigma_3^2\,,
\end{align}
where $f(\tau), g(\tau)$ are two scale factors (which are straightforwardly related to $a(t),\beta_+(t),$ see below), $l$ is a convenient parameter that will be related to the vacuum energy, and the time coordinate is re-scaled. In these coordinates, the action is given by
\begin{align}
S = Vol_3 \int d\tau \left[ \frac{1}{8l^2\sigma} \left[ 16l^4 - 2\sigma^2 {f}_{,\tau}{g}_{,\tau} - \frac{f}{g}\left(4l^4 + \sigma^2 {g}_{,\tau}^2\right) \right] - \frac{2l^2}{\sigma}g \Lambda \right]\,.
\end{align}
Varying with the respect to the fields gives the following equations of motion
\begin{align}
g{g}_{,\tau\tau} - \frac{1}{2}{g}_{,\tau}^2 - 2\frac{l^4}{\sigma^2} = 0\,, \\
{f}_{,\tau\tau} + \frac{{g}_{,\tau}}{g}{f}_{,\tau} + \frac{4l^4}{\sigma^2}\frac{f}{g^2} - \frac{8l^4}{\sigma^2} \Lambda = 0\,.
\end{align}
Now we can state the AO solution, which is given by \cite{Anabalon:2018rzq}
\begin{align}
g(\tau) &= \frac{l^2}{\sigma}\left(\tau^2 +1 \right)\,, \\
f(\tau) &= \frac{4l^2}{\sigma^2}\frac{\tau^4 + (6 - \sigma)\tau^2 + \mu \tau + \sigma - 3}{\tau^2 +1}\,,
\end{align}
supplemented with the identification of $l$ as being the Hubble length, $\Lambda = \frac{3}{l^2}$. The solution contains two free parameters, namely $\sigma$ and $\mu.$ We see that $g(\tau)$ is manifestly positive but having $f(\tau)$ non-zero everywhere imposes the following conditions on the free parameters $\sigma$ and $\mu,$
\begin{align}
3 < \sigma < 12\,, \qquad |\mu| < \frac{2}{9}\sqrt{3} \sqrt{\sigma - 3}(12 - \sigma)\,.
\end{align}
To better illustrate the physical meaning of the variables, we translate them into the scale factor $a$ and anisotropy $\beta_+$. By comparing the metrics we find that
\begin{align}
a = \left(f g^2 \right)^{1/6}\,, \qquad \beta_+ = \frac{1}{3} \ln \frac{g}{f}\,. \label{dictionary}
\end{align}
The isotropic (de Sitter) limit is restored in the case $\mu = 0$ and $\sigma = 4$. The parameters are related to the anisotropy and its derivative at $\tau = 0;$ in fact we have 
\begin{align}
\beta_+(0) &= \frac{1}{3} \ln \frac{\sigma}{4(\sigma - 3)}\,, \label{betatb}\\
\frac{d\beta_+}{d\tau}(0) &=  - \frac{1}{3(\sigma -3)} \mu\,, \label{betadotatb}
\end{align}
while asymptotically we have 
\begin{align}
\beta_+ (\pm \infty) &= \frac{1}{3} \ln \frac{\sigma}{4} + {\cal{O}}(\tau^{-2})\,, \label{betainf}\\
\frac{d\beta_+}{d\tau}(\pm \infty) &=  0 + {\cal{O}}(\tau^{-3})\,. \label{betadotinf}
\end{align}
Eq. \eqref{betadotatb} shows that $\mu$ determines the velocity of the anisotropy at the bounce, and as a consequence also the amount of time asymmetry of the solution. Meanwhile Eq. \eqref{betatb} implies that non-singular bounces can occur for all values of $ \beta_+ > \frac{1}{3}\ln\left(\frac{1}{3}\right) \approx -0.366$ at the bounce. In particular, the anisotropy can be arbitrarily large in the positive $\beta_+$ direction, as expected from the shape of the potential. However, there is a lower limit at $\frac{1}{3}\ln\left(\frac{1}{3}\right),$ which is not at the point where the potential turns positive, but rather a little into the negative potential region. This is illustrated in the left panel of Fig.~\ref{fig:bounceaxis}. This limiting value also agrees with our numerical results, cf. the location of the re-collapse region on the $\beta_-=0$ axis in Fig.~\ref{fig:bouncenumber}. Asymptotically, Eqs. \eqref{betainf} and \eqref{betadotinf} imply that the anisotropy parameter tends to a constant value, and this value is forced to be in the rather small range $\frac{1}{3}\ln\left( \frac{3}{4}\right) < \beta_+(\infty) < \frac{1}{3}\ln 3.$ This range reflects the focussing effect towards small values of the anisotropy that we already discussed in subsection \ref{sec:sym}.

The exact solution permits us to understand a few additional features analytically. From Eq. \eqref{dictionary} we can see that $a^6$ is a 6th order polynomial in time, implying that it can have $5$ extrema at most. Explicitly, we have
\begin{align}
a^6 = \frac{4l^6}{\sigma^4}\left[ \tau^6 + (7-\sigma) \tau^4 + \mu \tau^3 + 3 \tau^2 + \mu \tau + \sigma - 3\right]\,.
\end{align}
For time symmetric solutions, with $\mu=0,$ the extrema are then given by the real solutions to the equation
\begin{align}
\tau \left( \tau^4 + \left( \frac{14}{3} - \frac{2\sigma}{3}\right)\tau^2 + 1\right) = 0\,.
\end{align}
This straightforwardly implies that time symmetric axial Bianchi IX bounces have a single minimum of the scale factor for $3 < \sigma < 10,$ a minimum and two inflection points for $\sigma = 10$ and $3$ bounces separated by $2$ local maxima for $10 < \sigma < 12.$ We cannot have more than $3$ bounces for these solutions, a feature that can be understood intuitively in the sense that the vanishing of $\beta_-$ implies that the potential only contains a wall in the negative $\beta_+$ direction, and multiple BKL-type reflections off the potential walls cannot occur. A similar calculation shows that the extrema of $\beta_+$ occur (again in the time symmetric $\mu=0$ case) when 
\begin{align}
\tau=0, \quad \text{and}\,\text{when} \quad \tau^2 = \frac{3\sigma- 12}{\sigma - 4}\,.
\end{align} 
Thus, except for the de Sitter solution at $\sigma=4,$ the anisotropy always has $3$ extrema, a fact that is also nicely seen in the right panel of Fig. \ref{fig:bounceaxis}.

In the appendix we show that the closely related Kantowski-Sachs metric, in which the spatial sections contain a two-sphere rather than a three-sphere, also admit non-singular bounce solutions that are easily describable by an analytic solution, and that have related properties.


\section{Bounces in the presence of a scalar field} \label{sec:scalar}

Up to now we modelled dark energy via a cosmological constant. However, we may also consider the possibility that dark energy evolves over time, a situation which can be described by using a scalar field  $\phi$ in a potential $V(\phi)$.  Then the equations of motion are augmented to include contributions from the (minimally coupled) scalar, to become
\begin{align}
 & \frac{\ddot{a}}{a} + \frac{1}{2}\left( \dot{\beta}^2_+ + \dot{\beta}^2_- \right)  -\frac{1}{3}\left( \dot{\phi}^2 - V(\phi) \right)  = 0\,, \label{eqscalara}\\
& \ddot{\beta}_+ + 3H\dot{\beta}_+  + \frac{2}{3 a^2} U_{,\beta_+} = 0\,, \\
&  \ddot{\beta}_- + 3H\dot{\beta}_-  + \frac{2}{3 a^2} U_{,\beta_-} = 0\,. \\
&  \ddot{\phi} + 3H\dot{\phi}  + V'(\phi) = 0\,. \label{phi}
\end{align}
while the Friedman equation becomes
\begin{align}
3H^2 = V(\phi)+\frac{1}{2}\dot{\phi}^2 + \frac{3}{4}(\dot{\beta}_+^2+\dot{\beta}_-^2) + \frac{1}{a^2}U(\beta_+, \beta_-)\,. \label{eqscalarF}
\end{align}
This setting is familiar from inflation and quintessence models of dark energy. There are some similarities here, as one of the conditions for obtaining a bounce is that the strong energy condition must be violated, and this is the same condition as that for accelerated expansion. The scalar field equation of state is given by the ratio of pressure to energy density, which in the cosmological context can be expressed as
\begin{align}
w= \frac{\frac{1}{2}\dot\phi^2 - V(\phi)}{\frac{1}{2}\dot\phi^2 + V(\phi)}\,.
\end{align}
A violation of the strong energy condition corresponds to $w < - \frac{1}{3}.$ In inflation and quintessence models, this condition is realised by the field slowly rolling down the potential, so that the kinetic energy is sufficiently small compared to the potential energy, more precisely such that  $\frac{1}{2}\dot\phi^2 < \frac{1}{2}V(\phi)$ (in the absence of anisotropies, Eq. \eqref{eqscalara} then immediately implies $\ddot{a}>0$). This regime where the scalar slowly rolls down the potential is required for such a phase to last for an extended period of time, and for this reason the potential must not only be sufficiently flat in one location, rather it must be so over an extended field range. For non-singular bounces, one could consider a similar scenario where the scalar field rolls down while the universe bounces. This works less well than for inflation/quintessence however, as the scalar field kinetic energy is blue-shifted during contraction, and thus the standard Hubble friction term in the equation of motion \eqref{phi} becomes an anti-friction term. There is however an alternative manner in which a scalar potential can usefully lead to a bounce, and this is to consider the situation in which the scalar field runs \emph{up} the potential during the contracting phase. It can do so again because of the blue-shifting. Moreover, one can then imagine the situation where the scalar slows down as it rolls up, comes to rest at (or around) the time of the bounce, and subsequently rolls down again during the expansion phase. A great advantage of this scenario is that once the scalar comes to rest, the equation of state is precisely that of a cosmological constant, $w=-1.$ And for a bounce, which occurs over a relatively short time scale, this is enough. One does not need an extended period of strong energy violation. This implies that bounces can occur even in potentials that are rather steep (in fact, one can momentarily achieve $w=-1$ in any potential), and that one would not consider for inflationary model building. That said, we should now look at the combination of all the conditions required for a bounce, and then compare to numerical examples.

\begin{figure}[h] 
\begin{center}
\includegraphics[width=0.45\textwidth]{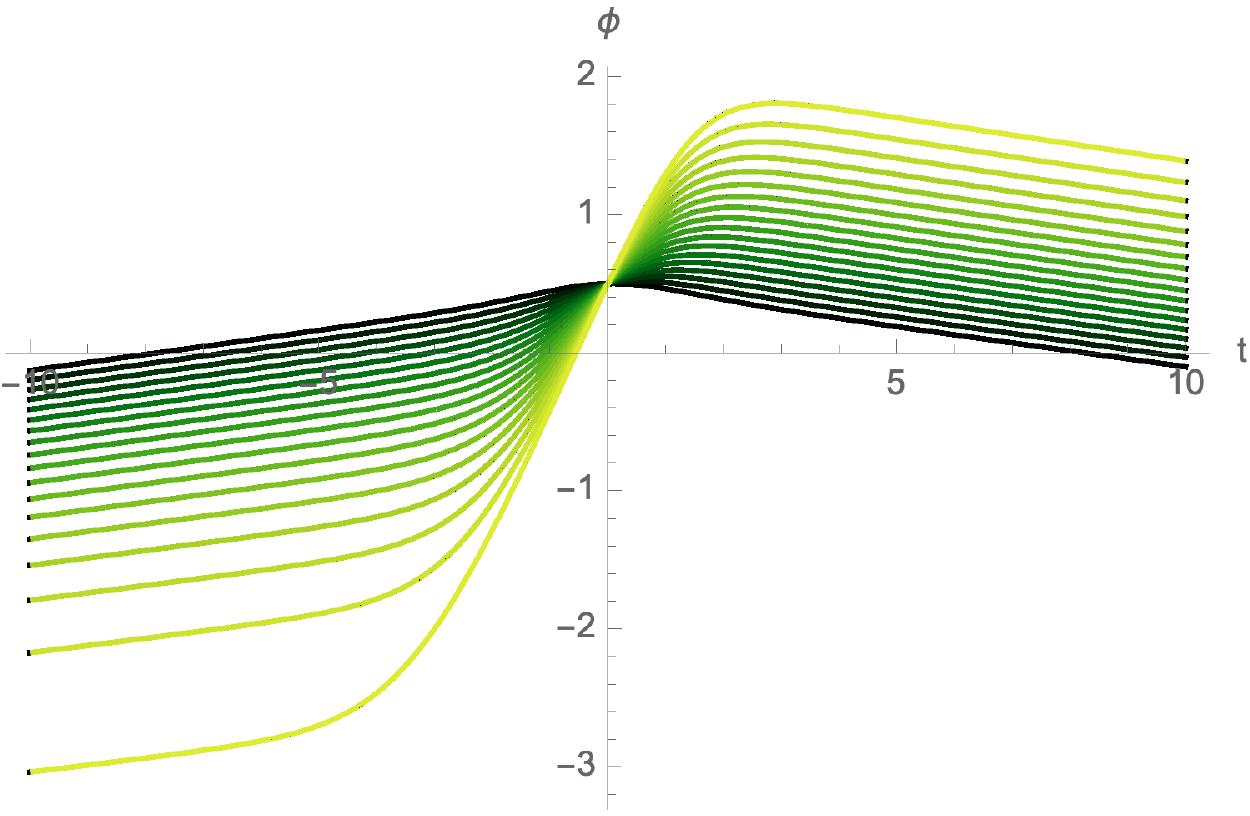}
\includegraphics[width=0.45\textwidth]{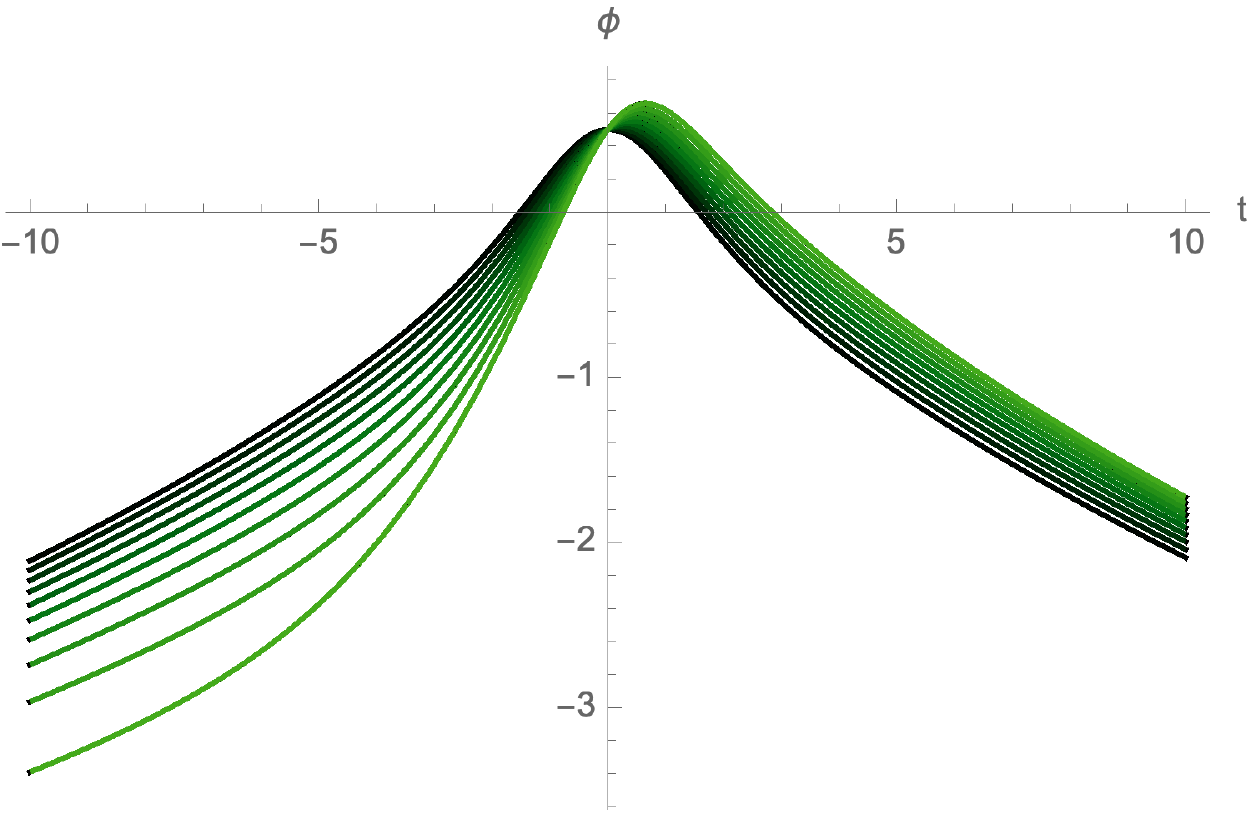}
\caption{These two plots show the evolution of the scalar field as a function of time, for solutions that bounce at $t=0$ at a specified value of $a=10,\phi=1/2,$ for a range of values of $\dot\phi$ at the bounce. The left panel is for a potential $V=e^{\phi/10},$ showing solutions with $\dot\phi(t=0)$ up to values of $0.90$ while the right panel is for $V=e^{\phi/2},$ showing solutions with $\dot\phi(t=0)$ up to values of $0.48$. Lighter curves correspond to larger velocities at the bounce.}
\label{fig:scalaratbounce}
\end{center}
\end{figure}

The minimal value of the scale factor at the bounce can again be found from the Friedman equation \eqref{eqscalarF}, and is given by
\begin{align}
a_b = \sqrt{\frac{-U}{V(\phi)+\frac{1}{2}\dot{\phi}^2 + \frac{3}{4}(\dot{\beta}_+^2+\dot{\beta}_-^2)}}\,.
\end{align}
The conditions to obtain a bounce are now given by 
\begin{align}
U(\beta_+, \beta_-) < 0 \quad \mid_{bounce}\,, \\
\frac{3}{2}(\dot{\beta}_+^2+\dot{\beta}_-^2) + \dot{\phi}^2 < V(\phi) \quad \mid_{bounce} \label{bouncecond2}\,.
\end{align}
We may again study a few numerical examples, this time for potentials of exponential form $V(\phi) = v_0 e^{c\phi}.$ We will also limit ourselves to cases with small anisotropies (in all the examples below we set $\beta_\pm= 1/100, \dot\beta_\pm = 0$ as an initial condition), since the inclusion of anisotropies is very similar to the discussions of the preceding sections. 

Fig. \ref{fig:scalaratbounce} shows the time evolution of the scalar field for a rather flat (left panel) and for a steeper potential (right panel). The initial conditions have been set at the bounce, which occurs at $t=0,$ for a range of values of the scalar field derivative. Thus the solutions that are plotted are automatically selected on the basis that a non-singular bounce occurs. As the scalar field derivative increases, the field runs further up the potential after the bounce, before eventually turning around and rolling back down. In potentials such as these, an inflationary phase would then follow. 

\begin{figure}[h] 
\begin{center}
\includegraphics[width=0.45\textwidth]{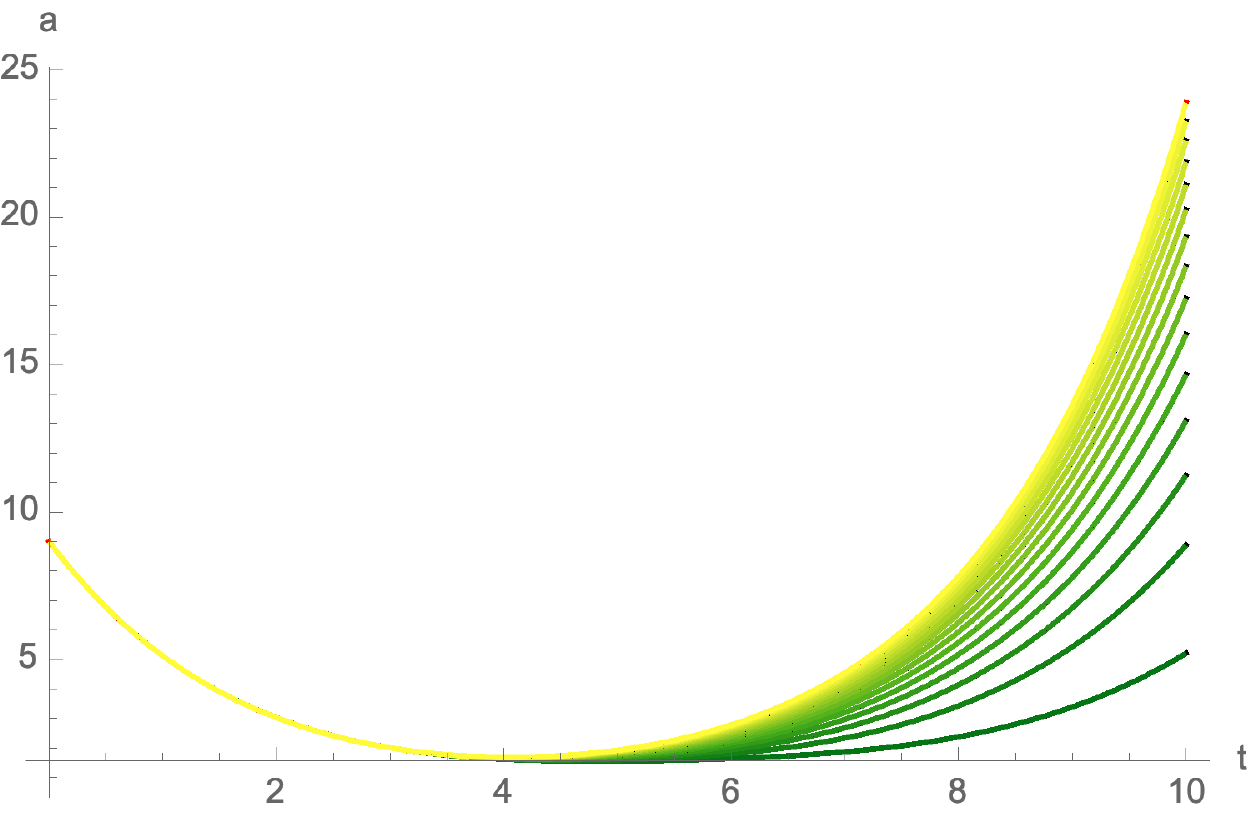}
\includegraphics[width=0.45\textwidth]{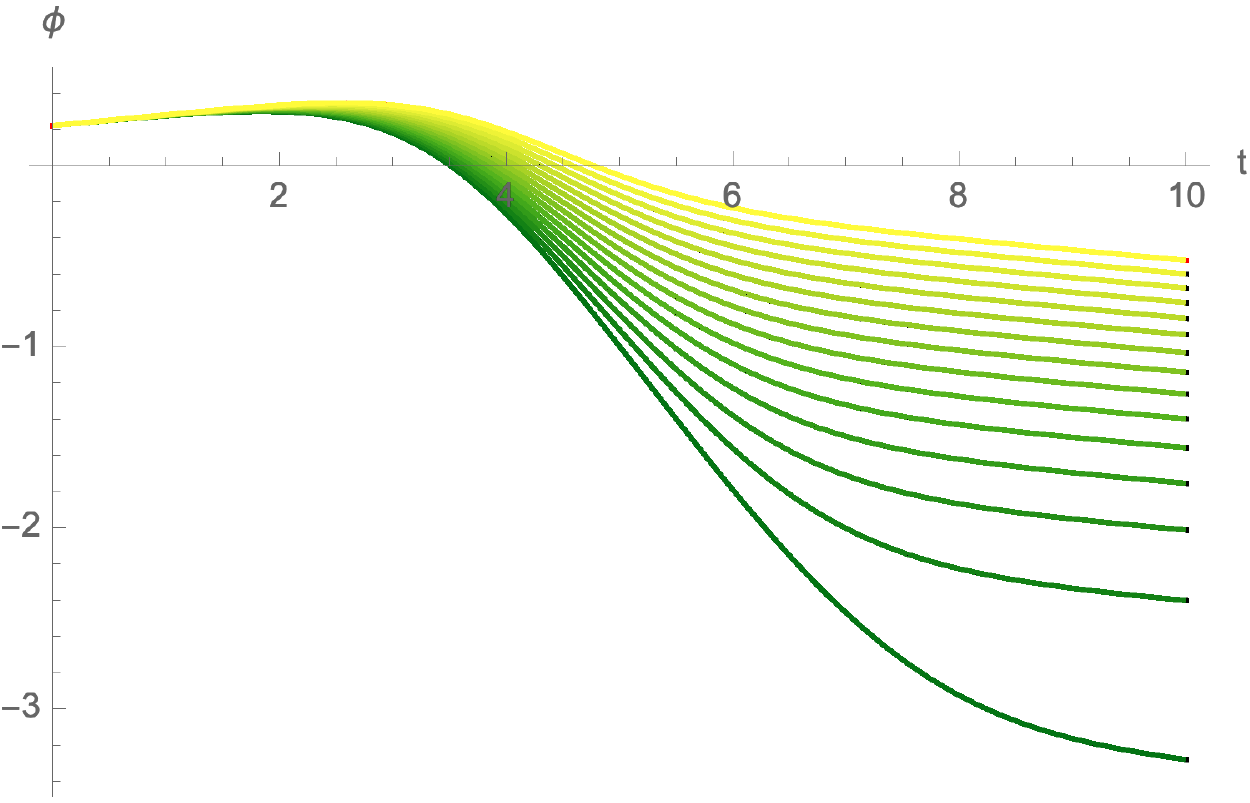}
\caption{The evolution of the scale factor and the scalar field for a range of initial velocities leading to non-singular bounce solutions. The potential is taken to be $\phi=e^{phi/10}$ here. The initial values for the scale factor and scalar field are $a_i=9, \phi_i=23/50,$ and the range of initial velocities leading to non-singular bounces is found to be $0.058 \lessapprox \dot\phi \lessapprox 0.064$, represented by the curves ranging from black to yellow respectively.}
\label{fig:scalarc10}
\end{center}
\end{figure}

\begin{figure}[h] 
\begin{center}
\includegraphics[width=0.45\textwidth]{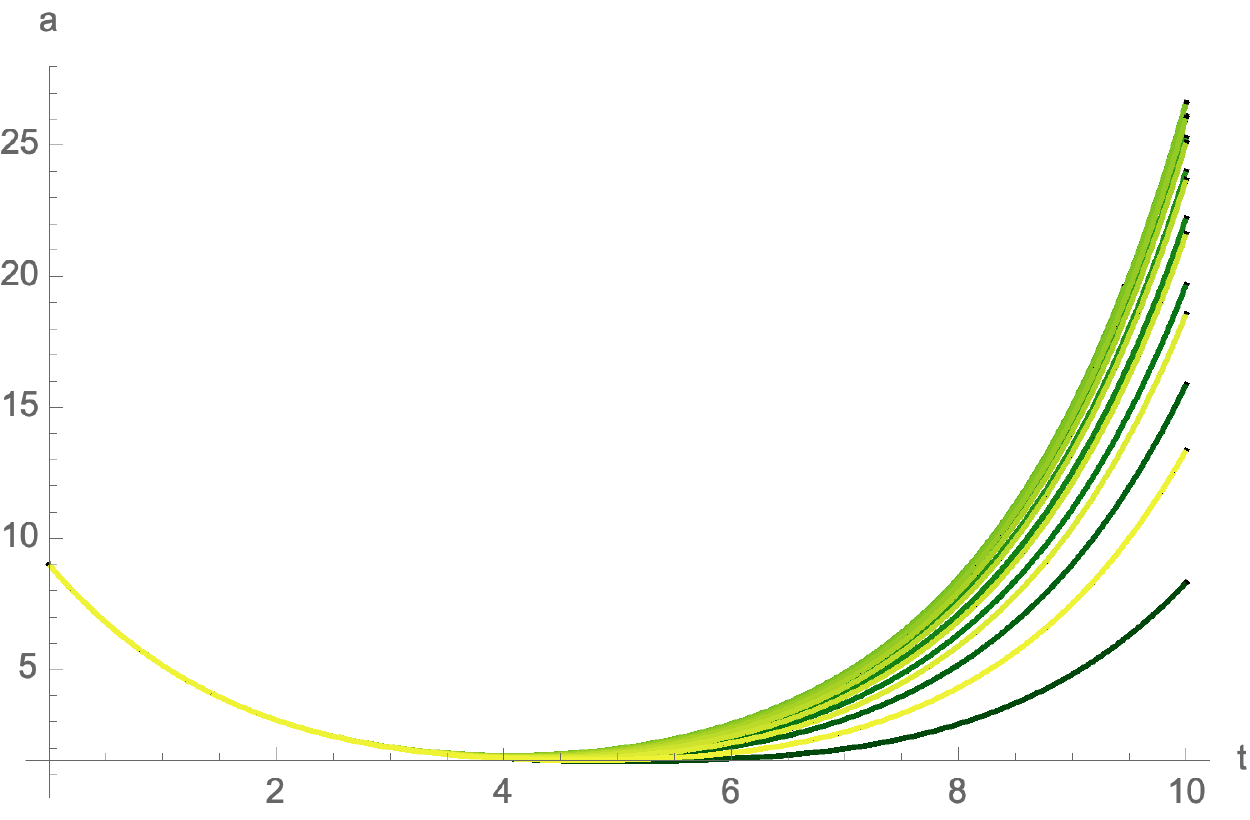}
\includegraphics[width=0.45\textwidth]{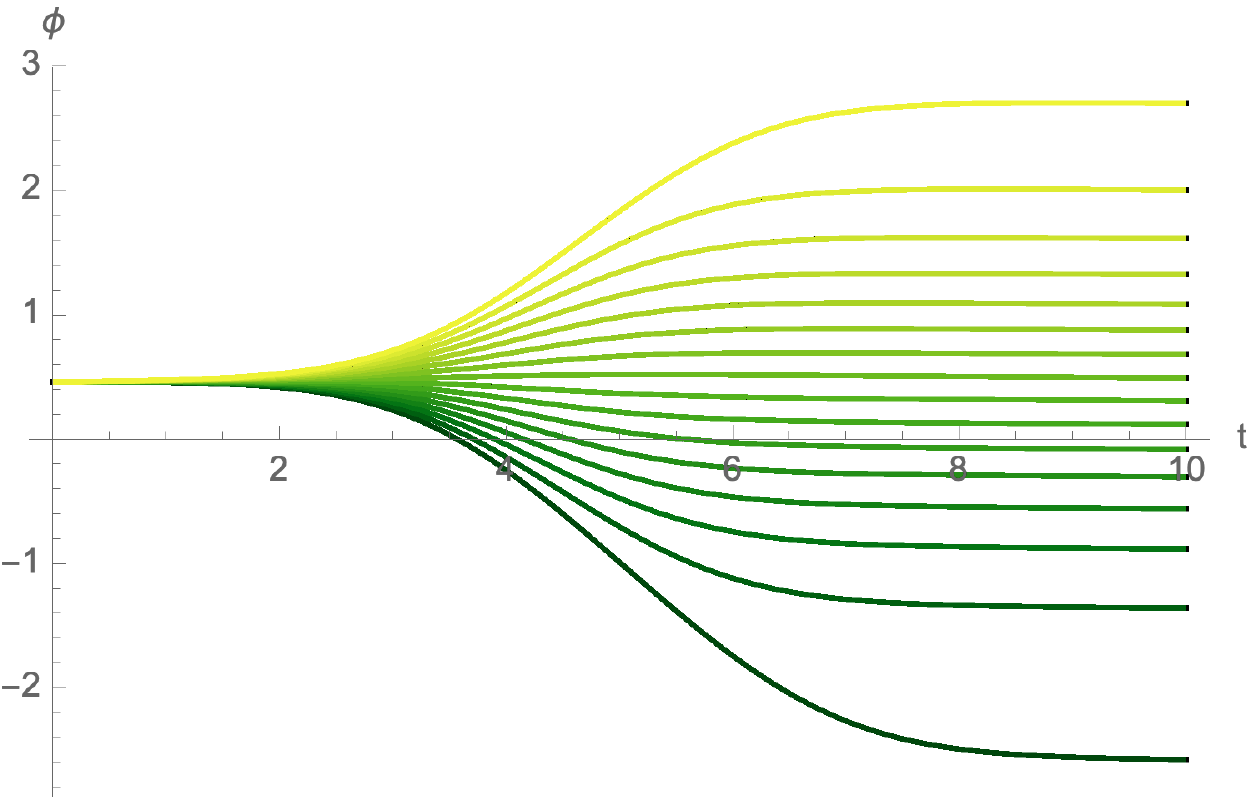}
\caption{Same as Fig. \ref{fig:scalarc10}, but for the potential $V=e^{\phi/100}$. The initial values for the scale factor and scalar field are $a_i=9, \phi_i=23/50,$ and the range of initial velocities leading to non-singular bounces is found to be $0.0024 \lessapprox \dot\phi \lessapprox 0.095$, represented by the curves ranging from black to yellow respectively.}
\label{fig:scalarc100}
\end{center}
\end{figure}

We may also set the initial conditions at an earlier time, in the contracting phase preceding a bounce (or a crunch). This is shown for two different potentials in Figs. \ref{fig:scalarc10} (with $V=e^{\phi/10}$) and \ref{fig:scalarc100} (with $V=e^{\phi/100}$). The solutions that are plotted are those that lead to non-singular bounces. For the steeper potential in Fig. \ref{fig:scalarc10}, we find that only solutions that first run up the potential lead to a bounce. In all these solutions the scalar field turns around at or shortly after the bounce, and rolls back down the potential. For the largest initial scalar field velocity, the field runs up the furthest, leading to the largest amount of expansion after the bounce. For a flatter potential, as shown in Fig. \ref{fig:scalarc100}, non-singular bounces may occur both when the scalar field runs up the potential, or down. Of course, eventually the field always rolls down the potential, and all these non-singular bounces are followed by phases of inflationary expansion. The largest amount of expansion right after the bounce occurs for the case where the scalar field velocity is practically zero at the bounce. For larger velocities, the bounce occurs somewhat later, so that there is less time for expansion. And for smaller initial velocities, the scalar rolls down the potential earlier, so that the bounce occurs while the field is already rolling down, implying a smaller expansion rate right after the bounce. In all cases, the range of initial velocities that lead to a non-singular bounce is small. This is mainly due to the blue-shifting of the scalar kinetic energy during the contraction phase, where one must ensure that the bound in Eq. \eqref{bouncecond2} does not get violated. We will discuss the initial conditions in more detail in the next section.


\section{Discussion} \label{sec:discussion}

In this paper, we have analysed non-singular bouncing universe solutions in the simplest possible setting: general relativity in the presence of a cosmological constant, or in the presence of a scalar field with a potential. Here bounces occur due to established physics only (dark energy is observationally established, as is the existence of a scalar particle, the Higgs), and without violations of the null energy condition. From this point of view, these bounces are considerably less speculative than bounces based on theories with specifically tuned higher-derivative kinetic terms, such as Galileon bounces \cite{Qiu:2011cy,Easson:2011zy,Ijjas:2016tpn}\footnote{Galileon models can be extended to supergravity theories \cite{Koehn:2013upa,Battarra:2014tga}, but it remains unclear whether they can arise in a truly fundamental framework, such as string theory \cite{Khoury:2012dn,Ovrut:2012wn}. Moreover, it is not clear if these theories are consistent at the quantum level, as they typically contain classes of unhealthy solutions in addition to the desired solutions.}. We have demonstrated that the bounces are robust under the inclusion of small anisotropies at the bounce, by studying bounce solutions within the Bianchi IX metric. Near the edge of the parameter space allowing for bounces, we have found solutions with multiple bounces, and accompanying turn-arounds of the anisotropy functions resulting from their non-linear dynamics. These multi-bounce solutions provide the link between non-singular single bounce solutions and chaotic BKL/mixmaster crunches that occur for ``most'' initial conditions.  

This brings us to the issue of initial conditions: even though the theory in which the bounces occur is very simple, the solutions themselves are very special \cite{Barrow:1980en}\footnote{Exactly how special depends on having a probability measure, an issue that is far from resolved in cosmology.}. A bounce occurs only if the energy density in homogeneous curvature, and that in dark energy, are larger at the time of the bounce than the kinetic energy coming from the time evolution of the anisotropies and of the scalar field. However, the kinetic energy in the scalar grows much faster during a contracting phase (neglecting the potential, it is proportional to $1/a^6$) than the homogeneous curvature ($\propto 1/a^2$) or the approximately constant dark energy density. Thus it remains an open problem as to what kind of dynamics during the contracting phase could lead to such non-singular bounces. An ekpyrotic phase cannot achieve this, as it suppresses curvature and leads to a fast-rolling scalar \cite{Khoury:2001wf,Lehners:2008vx}. We must leave this as an open question for the future. Let us just mention one mitigating thought: if the dark energy resides at a very high scale in the early universe, say very close to the Planck scale, then the range of allowed kinetic energies compatible with a bounce are rather large. In such a case, one would in fact only trust the theory (considered as an effective theory) for kinetic energies below the Planck scale, and thus the self-consistency of the assumptions would render the bounces more prevalent in the space of available solutions. 

If, in light of the preceding discussion, we simply assume that the conditions required for a non-singular bounce have been established, then the bounce would quite naturally lead into an inflationary phase afterwards. The bounce would imply that it is rather natural to find the scalar field high up on the potential (since it would have rolled up during the contracting phase), and the universe would have small anisotropies and be dominated by vacuum energy.  Furthermore, the inflationary phase would subsequently dilute the spatial curvature required for the bounce. In this sense, a non-singular bounce can provide a viable prelude to inflation (see also \cite{Piao:2003zm,Qiu:2015nha} for works in that direction). Of course, the question of initial conditions for inflation is then not solved, but shifted to the question of initial conditions for the bounce. One could then hope that this new viewpoint might lead to new ideas on how to address this open issue. For instance, could the bounce act as a kind of filter, thereby automatically selecting for universes with suitable ``initial'' conditions, similar in spirit to the scenario proposed for the ``phoenix'' universe \cite{Lehners:2008qe,Lehners:2009eg,Lehners:2010ug}?

There has been a renewed interest recently in cosmological models arising from string theory, due to proposed consistency requirements for string theoretic solutions \cite{Obied:2018sgi,Danielsson:2018ztv,Andriot:2018wzk}. These ``swampland'' criteria have put a lot of pressure on existing cosmological models, both of inflationary \cite{Agrawal:2018own,Garg:2018reu} and ekpyrotic \cite{Lehners:2018vgi} type, as they suggest in particular that the scalar field range $\Delta\phi$ must remain smaller than order one in Planck units in a consistent effective description, while any positive potential must remain sufficiently steep throughout ($|V'|/V$ larger than some order one number in Planck units). In the present context it is interesting to point out that non-singular bounces can easily fulfill these criteria: given that the scalar field can run up the potential and come back down afterwards, the range of field values that is traversed can naturally remain small. Moreover, it is not necessary that the scalar field potential be flat: as the scalar comes to rest on the potential, it momentarily acts like a cosmological constant, regardless of the steepness of the potential, and, as we have seen, this can be sufficient to induce a cosmological bounce.

Apart from the open questions listed so far, there are two further avenues for future research that seem particularly promising: the first is related to the question as to what happens when the anisotropy becomes larger than the allowed bound at the bounce, i.e. what happens when the anisotropy potential $U(\beta_+,\beta_-)$ becomes positive? Here, no classical non-singular bounce solutions remain, but perhaps there exist quantum transitions between a contracting and an expanding phase of the universe. In the absence of anisotropies, we have started an exploration of such solutions in \cite{Bramberger:2017cgf}, but it would be important to extend this analysis to the more rigorous approach using Picard-Lefschetz theory to define and evaluate the gravitational path integral \cite{Feldbrugge:2017kzv,Feldbrugge:2017mbc}, while also including anisotropies. 

Finally, as emphasised by Anabal\'{o}n and Oliva \cite{Anabalon:2018rzq}, bouncing universe solutions with positive vacuum energy are closely related to wormholes in the presence of negative vacuum energy. Hence our results suggest the existence of many new anisotropic wormhole solutions, including multi-wormhole solutions with arbitrarily large numbers of throats. It will be interesting to construct and study these solutions, which we hope to do in the near future.


\acknowledgments

We would like to thank Andr\'{e}s Anabal\'{o}n for discussions and comments on the manuscript, and Job Feldbrugge and Neil Turok for stimulating discussions that led to the idea for this project. We gratefully acknowledge the support of the European Research Council in the form of the ERC Consolidator Grant CoG 772295 ``Cosmology'' and the Studienstiftung des Deutschen Volkes.


\appendix

\section{Kantowski-Sachs bounces} \label{sec:KS}

An easier toy model for non-singular cosmological bounces than the axial Bianchi IX model of section \ref{sec:exact} can be found in the Kantowski-Sachs (KS) class of metrics \cite{Kantowski:1966te}. These metrics contain a two-sphere in their spatial directions, and the line element is given by
\begin{align}
ds_{KS}^2 = - dt^2 +\frac{a^2(t)}{4}e^{-2\beta(t)} dr^2 + \frac{a^2(t)}{4}e^{\beta(t)} d\Omega_2^2 \,,
\end{align}
where the factor of $1/4$ was included in analogy with the Bianchi IX case. Again, $a$ represents the spatial volume while $\beta$ quantifies an anisotropic deformation. In this case there is only one a deformation parameter. In the presence of a cosmological constant $\Lambda,$ the equations of motion and constraint are given by
\begin{align}
3\frac{\ddot{a}}{a} + \frac{3}{2}\dot\beta^2 &=N^2 \Lambda \\
\ddot\beta+ 3H\dot\beta + \frac{2N^2}{3a^2}U_{,\beta} &= 0 \\
3H^2 &= \frac{3}{4}\dot\beta^2 + N^2 (\Lambda + \frac{1}{a^2}U)
\end{align}
where the constraint has been used to simplify the acceleration equation. The effective potential is
\begin{align}
U(\beta) = -4 e^{-\beta}\,.
\end{align}
It is very similar to the axial Bianchi IX potential in Eq. \ref{anisotropycornerpotential}, except that the $e^{-4\beta}$ term is absent. At large positive $\beta$ the two models are essentially equivalent, but at negative $\beta$ the KS potential remains negative, causing a runaway of the solutions asymptotically.

\begin{figure}[h] 
\begin{center}
\includegraphics[width=0.4\textwidth]{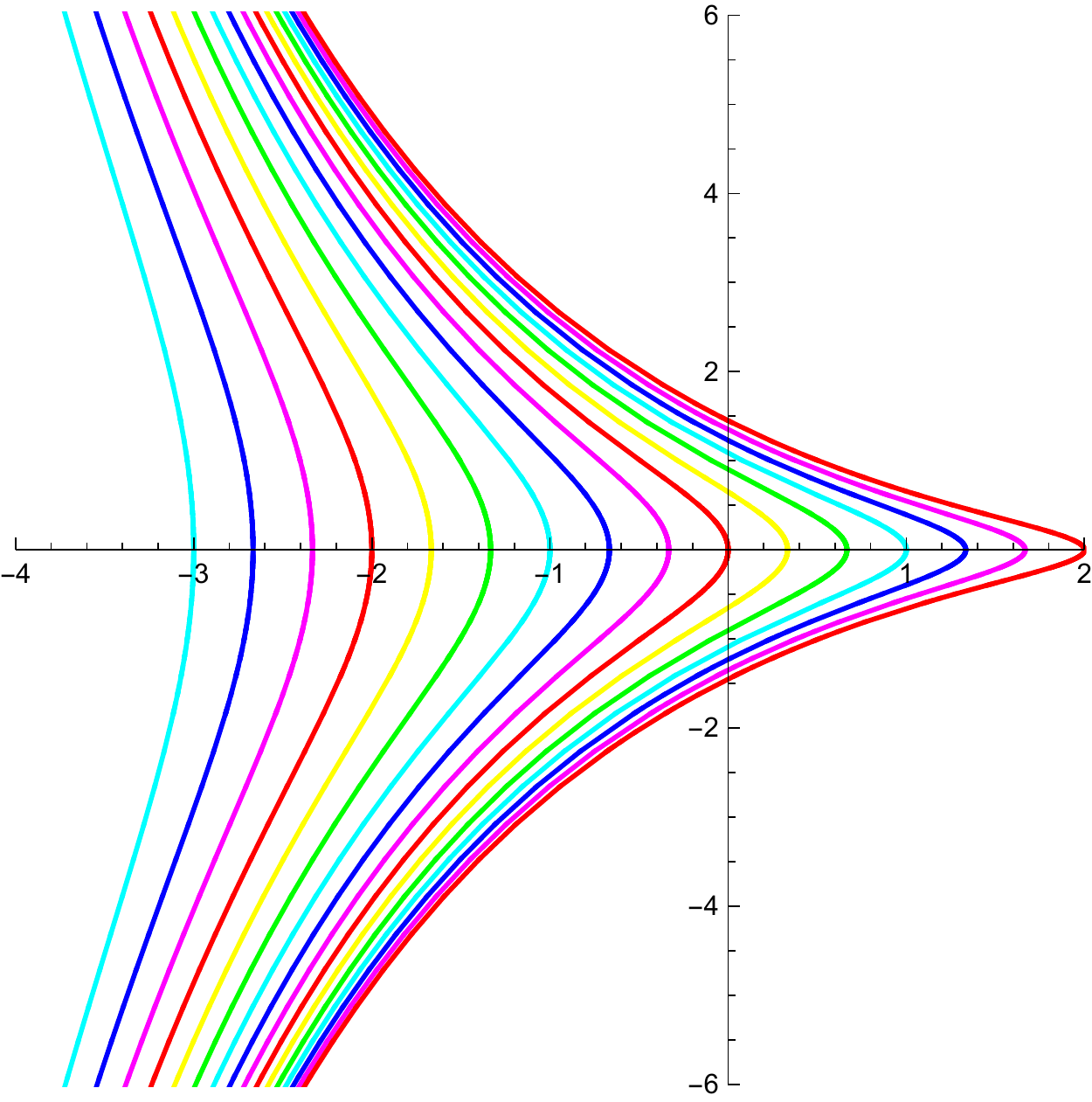}
\caption{Evolution of the anisotropy parameter $\beta$ as a function of time for Kantowski-Sachs bounces. For all $\beta$s there exists a bounce (here with $\dot\beta=0$ at the bounce). }
\label{fig:KSbounces}
\end{center}
\end{figure}

Now we may look for actual bounce solutions. A perturbative expansion around a would-be bounce leads to the expansions
\begin{align}
a &= a_b (1+ \frac{\Lambda}{6}t^2 + \cdots) \\
\beta &= \beta(0) - \frac{1}{3}\Lambda t^2 + \cdots
\end{align}
where the scale factor at the bounce is given by $a_b = a(t=0)= \frac{1}{\sqrt{\Lambda}e^{\beta(0)/2}}$ and we have fixed the time of the bounce to be at $t=0.$ The above expansions suggest that one might try an ansatz $a \propto e^{-\beta/2}$ and this indeed solves the equations of motion exactly,
\begin{align}
a &= \frac{1}{\sqrt{\Lambda}} e^{-\beta/2} = c_1 \left(\cosh(\sqrt{\Lambda} t + c_2)\right)^{1/3}
\end{align}
where $c_1,c_2$ are integration constants. Thus analytic bounce solutions exist for every possible value of $\beta$ at the bounce, while asymptotically the anisotropy parameter $\beta$ always runs off to minus infinity. For these solutions, in fact only the $r$ direction bounces while the 2-sphere remains constant throughout. These solutions are plotted in Fig. \ref{fig:KSbounces}, and may be recognised as $dS_2 \times S^2$ (and we note that closely related wormhole solutions also exist \cite{Anabalon:2012tu}).

\bibliographystyle{utphys}
\bibliography{NonSingularBounces}

\end{document}